\documentclass[journal]{IEEEtran}
\usepackage{amsmath}
\usepackage{amssymb}
\usepackage{amsthm}
\usepackage{cite}
\usepackage{color}
\usepackage{xcolor}
\usepackage{epsfig,latexsym}
\usepackage{float}
\usepackage{fancyhdr}
\usepackage{graphicx}
\usepackage{indentfirst}
\usepackage{mdwmath}
\usepackage{mdwtab}
\usepackage{subfigure}
\usepackage{setspace}
\usepackage{times}
\usepackage{url}
\usepackage{multirow}
\usepackage{algorithm}
\usepackage{algorithmic}
\usepackage{enumitem} 

\newtheorem{remark}{Remark}
\newtheorem{theorem}{Theorem}

\newtheorem{lemma}{Lemma}

\newtheorem{corollary}{Corollary}

\begin{document}

\title{Rate Maximization for Multi-Waveguide PASS:\\
 A Hierarchical User Scheduling and\\
  Joint Optimization Framework}
\author{Guangyu Li,
        Xin Sun,
        Tianwei Hou,~\IEEEmembership{Member,~IEEE,}
        Anna Li,~\IEEEmembership{Member,~IEEE,}\\
        Yuanwei Liu,~\IEEEmembership{Fellow,~IEEE,}
        Arumugam Nallanathan,~\IEEEmembership{Fellow,~IEEE}
        
\thanks{ This work was supported in part by the Beijing Natural Science Foundation L232041, and in part by the EPSRC grant numbers to acknowledge are EP/W004100/1, EP/W034786/1, EP/Y037243/1 and EP/X04047X/2. (Corresponding author: Tianwei Hou.)}
\thanks{Guangyu Li, Xin Sun and Tianwei Hou are with the School of Electronic and Information Engineering, Beijing Jiaotong University, Beijing 100044, China (e-mail: 22331095@bjtu.edu.cn, xsun@bjtu.edu.cn, twhou@bjtu.edu.cn). }
\thanks{Anna Li is with the School of Computing and Communications, Lancaster University, Lancaster LA1 4WA, U.K. (e-mail: a.li16@lancaster.ac.uk). }
\thanks{Yuanwei Liu is with the Department of Electrical and Electronic Engineering, The University of Hong Kong, Hong Kong (e-mail: yuanwei@hku.hk). }
\thanks{Arumugam Nallanathan is with the School of Electronic Engineering and Computer Science, Queen Mary University of London, London E1 4NS, U.K., and also with the Department of Electronic Engineering, Kyung Hee University, Yongin-si, Gyeonggi-do 17104, Korea (e-mail: a.nallanathan@qmul.ac.uk).}

}
\maketitle

\begin{abstract}
Pinching-antenna systems (PASS) have emerged as a promising flexible-antenna architecture capable of dynamically reconfiguring wireless channels by activating dielectric particles along waveguides. The sum rate maximization problem in multi-waveguide PASS is investigated in this study. Both in-waveguide propagation loss and coupling effects are explicitly modeled. To tackle the optimization problem, a hierarchical user scheduling (HUS) algorithm is proposed. The HUS algorithm minimizes the sum of squared distances between users and their associated waveguides to mitigate path loss. Additionally, spatially separated users are assigned within each time slot to reduce inter-user interference. Furthermore, a joint optimization framework integrating power allocation and pinching-antenna (PA) positioning is developed to further improve system sum rate. Specifically, PAs' positions are optimized via one-dimensional search, while the power allocation problem is solved by using the Lagrangian duality and fractional programming. Numerical results show that the HUS algorithm clearly outperforms random pairing, and the proposed power allocation algorithm shows a marked performance improvement over the maximum ratio transmission algorithm. Moreover, the results explicitly demonstrate the considerable impact of in-waveguide propagation loss and coupling effects on the performance of PASS. 
\end{abstract}


\section{Introduction}

As the demand for flexibility in next-generation (NG) wireless networks becomes essential, the conventional fixed antennas fall short in meeting the rising performance demands \cite{6375940,larsson2014massive,8869705}. Flexible-antenna systems, which allow dynamic repositioning of radiating elements, are capable of reshaping wireless channels in real time \cite{ahmadzadeh2025enhancedovertheairfederatedlearning}. As a result of introducing additional spatial degrees of freedom, the system can provide additional channel gains \cite{10437926,11100909}. 

Typical flexible-antenna systems include reconfigurable intelligent surface (RIS) and movable antenna (MA) systems. RIS enables precise control of reflected electromagnetic (EM) waves, thereby reconstructing the wireless channel \cite{9424177}. In contrast, MA technology enables local antenna movement within a confined region to exploit spatial degrees of freedom \cite{10909572}. Although these technologies offer significant potentials for performance enhancement \cite{9427474,10643473,9140329}, they face critical limitations. RIS suffers from severe path loss over large discrete distances \cite{9690635}. For MA systems, the channel state information acquisition over the movement region may incur prohibitive estimation overhead \cite{10286328}. As an innovative flexible antenna technology, the pinching-antenna systems (PASS) have attracted increasing attention from researchers because of its effective application in waveguide systems.
\vspace{-0.9mm}
\subsection{Related Works}
PASS were first introduced by NTT DOCOMO as a novel approach to enhance the performance in wireless communication systems \cite{suzuki2022pinching}. In contrast to traditional flexible antenna systems, PASS utilize dielectric waveguides as the transmission medium, along which several radiating elements are enabled by small and cost-effective dielectric particles \cite{yang2025pinchingantennasprinciplesapplications}. These elements, known as pinching-antennas (PAs), can be dynamically activated, deactivated, or relocated along the waveguide, enabling the system to reconfigure both large-scale and small-scale channel conditions \cite{10909665}. The flexible architecture effectively reduces path loss and supports the line-of-sight (LoS) links, both critical for NG wireless networks \cite{liu2025pinching,liu2025pinchingantennasystemspasstutorial,ding2024flexibleantennasystemspinchingantennaperspective}.  PASS facilitate a novel beamforming paradigm known as pinching beamforming, which simultaneously adjusts both the path loss and phases experienced by the radiated signals \cite{xu2025jointtransmitpinchingbeamforming}. The capability to adjust these parameters leads to efficient signal enhancement and interference mitigation \cite{zhao2025waveguidedivisionmultipleaccess,wang2025antennaactivationresourceallocation}. 

Under the ideal waveguide condition with negligible propagation loss, the theoretical performance bounds of PASS were extensively analyzed. A closed-form upper bound on the array gain was derived in \cite{ouyang2025arraygainpinchingantennasystems}, where the optimal number and spacing of PAs were derived to govern the performance ceiling. The work in \cite{wang2024antennaactivationnomaassisted} investigated the theoretical throughput benchmarks of non-orthogonal multiple access (NOMA) assisted PASS and demonstrated its potential performance superiority over conventional fixed-antenna systems. The study in \cite{11195810} analyzed the ergodic rate of uplink PASS, and showed that optimizing the PAs’ positions can improve the ergodic sum rate. The study in \cite{zhang2025uplinksumratemaximization} optimized the resource allocation for uplink multiuser multiple-input single-output (MISO) PASS, and demonstrated that PASS can achieve a higher uplink sum rate compared to conventional fixed-antenna designs. Furthermore, the work in \cite{zhao2025pinchingantennasystemsenabledmultiusercommunications} analyzed the max-min fairness bounds of multi-waveguide PASS under different transmission structures, highlighting its performance advantage over conventional multiple-input multiple-output (MIMO) systems. In summary, these studies under ideal waveguide conditions established the theoretical performance bounds of PASS, serving as benchmarks for subsequent work.


In practice, in-waveguide propagation loss became a dominant impairment, particularly for high-frequency applications such as millimeter-wave and terahertz communications. Unlike idealized models that assumed negligible loss, recent works had explicitly incorporated in-waveguide attenuation into the system design and optimization. The study in \cite{tyrovolas2025performanceanalysispinchingantennasystems} analyzed the outage probability and average rate of PASS under in-waveguide propagation losses, showing their significant impact on system performance. Subsequent work in \cite{hu2025sumratemaximizationpinchingantennaassisted} advanced the model by explicitly incorporating in-waveguide attenuation into the link budget and signal strength analysis, revealing that in-waveguide loss constrained the activation distance and optimal placement of PAs. To further enhance modeling accuracy, the study in \cite{li2025pinchingantennaaidedwirelesspowered} developed a joint optimization framework that simultaneously configured PAs' positions and beamforming weights, with in-waveguide loss formulated as a function of propagation distance and carrier frequency. Most recently, the study in \cite{sun2025multiuserbeamformingpinchingantennasystems} derived closed-form and near-closed-form solutions that embed in-waveguide attenuation into the achievable rate expressions and system sum rate, enabling tractable performance bounds under lossy conditions.

Another important factor that affected the performance of PASS was the coupling effect between waveguides and PAs. The phenomenon became particularly significant when multiple PAs were activated along a dielectric waveguide, leading to mutual interactions that could distort the intended radiation pattern. The coupling effect was first explicitly modeled in \cite{wang2025modelingbeamformingoptimizationpinchingantenna}, where a physical model was established to quantify the coupling-induced impedance and radiation changes as a function of spatial arrangement of PAs along the waveguide. Building upon this foundation, the study in \cite{xu2025pinchingantennasystemspasspower} revealed how coupling modified both the amplitude and phase of radiated signal. This enabled more accurate modeling of PASS deployments, which highlighted the need to optimize the activation of PAs to mitigate coupling-induced performance degradation.

\vspace{-2.9mm}
\subsection{Motivation and Contribution}

Existing state-of-the-art studies on PASS have developed along three largely parallel lines of research, i.e., idealized performance analysis, lossy waveguide aware modeling and design, and coupling aware electromagnetic characterization. Nevertheless, the current literature still falls short in several key aspects. In particular, there is still no multi-waveguide PASS model that jointly characterizes both in-waveguide propagation loss and waveguide–PA coupling effects. Moreover, the joint sum rate maximization problem with PAs' positions optimization and precoding design, under users’ minimum rate constraints, remains largely unexplored. In addition, user scheduling, which can further improve system performance by exploiting temporal and spatial resources, has not been systematically studied in multi-waveguide PASS systems.


Motivated by these gaps, the main theoretical contributions of this paper are summarized below:
\begin{itemize}
\item We develop a unified physical model for multi-waveguide PASS that jointly captures in-waveguide propagation loss and waveguide-PA coupling effect, derived from EM field theory and coupled-mode theory. The model enables a more accurate characterization of the individual and joint impacts of the physical effects on the radiated signals and system performance.
\item We propose a hierarchical user scheduling (HUS) algorithm for the considered multi-waveguide PASS architecture. The algorithm minimizes the sum of squared distances to reduce path loss, and schedules spatially separated users in each time slot to mitigate interference. By improving the received signal-to-interference-plus-noise ratio (SINR) of all users, the HUS algorithm achieves a higher and more robust sum rate in a low-complexity manner.
\item We develop an alternating optimization (AO)-based framework for multi-waveguide PASS, which jointly optimizes PA positions and power allocation across waveguides to maximize the system sum rate under users’ minimum rate constraints. PAs' positions are optimized through one-dimensional search. The power allocation to each waveguide is optimized by using the Lagrangian duality and fractional programming (FP). The joint optimization framework leads to noticeable performance improvement in multi-waveguide PASS.
\item Numerical simulations are conducted to evaluate the impact of in-waveguide propagation loss and coupling effect, as well as the effectiveness of the proposed HUS algorithm and joint optimization framework. Our results demonstrate that: 1) In-waveguide propagation loss and coupling effects exert a pronounced influence on sum rate, underscoring the necessity of accurate EM modeling. 2) In multi-waveguide transmission, the coupling effect leads to discrete PA positioning, which enables flexible power redistribution and improves performance. 3) Optimized resource allocation and user scheduling provide greater benefits than simply increasing transmit power.
\end{itemize}

The remainder of this paper is organized as follows. Section II presents the system modeling of PASS, with an emphasis on the physical modeling of in-waveguide propagation loss and coupling effect. Section III focuses on the sum rate maximization of multi-waveguide PASS systems through HUS algorithm and joint optimization of power allocation and PA positioning. Section IV presents simulation results to validate the proposed algorithm. Section V concludes this paper.

\section{System Model}

\subsection{Antenna and Channel Model}

\begin{figure}[t]
    \centering
    \includegraphics[width=0.48\textwidth]{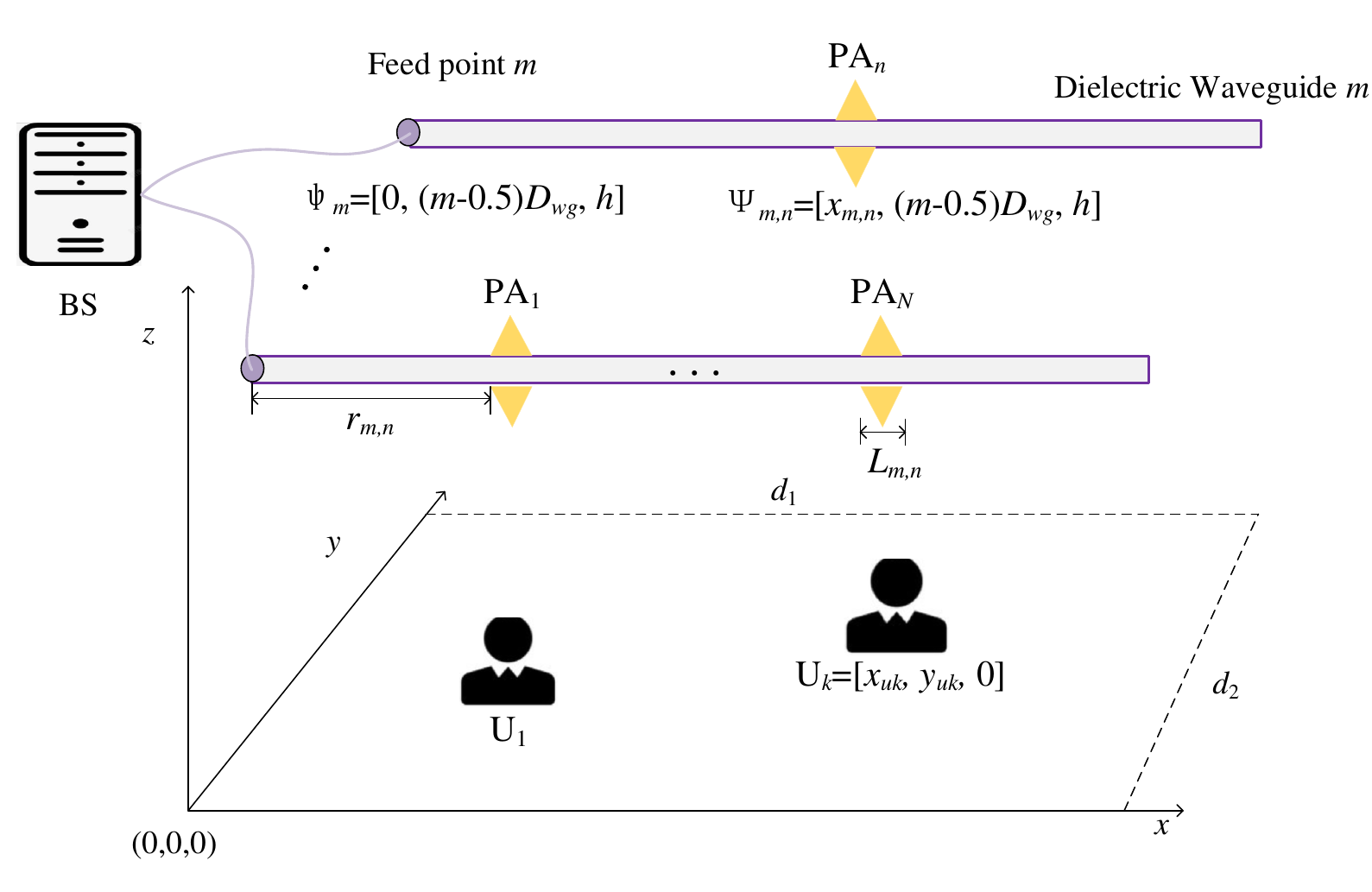} 
    \caption{Downlink pinching-antenna system model.} 
    \label{system} 
\end{figure}

As shown in Fig. \ref{system}, we consider a multi-waveguide downlink PASS serving $\textit{K}$ users. Users are uniformly distributed within a rectangular region with dimensions ${d_1} \times {d_2}$. $\textit{M}$ dielectric waveguides are located at a distance $\textit{h}$ above the $x$-$y$ plane. All waveguides are arranged with equal spacing in the $y$-axis direction with distance ${D_{wg}}$. The signal is fed from the base station (BS) to the feed point of each waveguide, which is denoted as ${\varphi _m} = \left[ {0,\frac{{\left( {2m - 1} \right){D_{wg}}}}{2},h} \right]$. The location of the  $\textit{m}$-th dielectric waveguide can be denoted as $ \left[ {\left( {0,D} \right),\frac{{\left( {2m - 1} \right){D_{wg}}}}{2},h} \right]$, where $\textit{D}$ is the dielectric waveguide length. 

$\textit{N}$ PAs are deployed on each dielectric waveguide to serve the users. In order to facilitate the system modeling, the position of the $\textit{k}$-th user is denoted as ${{\bf{U}}_k} = \left[ {{x_{uk}},{y_{uk}},0} \right]$. The position of the $\textit{n}$-th PA on the $\textit{m}$-th waveguide, i.e., the $(m,n)$-th PA, is denoted as ${\psi _{m,n}} = \left[ {{x_{m,n}},\frac{{\left( {2m - 1} \right){D_{wg}}}}{2},h} \right]$, where ${x_{m,n}}$ denotes the $x$-coordinate of its right endpoint. Let $L_{m,n}$ denotes the length of the $(m,n)$-th PA. Then, the $x$-coordinate of its left endpoint is denoted by $r_{m,n}$, satisfying $x_{m,n} = r_{m,n} + L_{m,n}$.


According to the spherical wave channel model\cite{9738442}, the channel vector between all PAs on the $\textit{m}$-th waveguide and the $\textit{k}$-th user is given by:
\begin{equation}\label{eq:los}
{{\bf{h}}_{m,k}} = {\left[ {\frac{{\eta {e^{ - j\frac{{2\pi }}{{{\lambda _0}}}{d_{k,m,1}}}}}}{{{d_{k,m,1}}}}, \cdots ,\frac{{\eta {e^{ - j\frac{{2\pi }}{{{\lambda _0}}}{d_{k,m,N}}}}}}{{{d_{k,m,N}}}}} \right]^T},
\end{equation}
where ${d_{k,m,n}}$ denotes the distance between the $\textit{k}$-th user and the $(m,n)$-th PA, $\eta  = \frac{c}{{4\pi {f}}} $, $c$ is the speed of light, and $f$ is the carrier frequency.

\vspace{-0.9mm}
\subsection{Physics-based Hardware Model} 
Existing coupling models for PASS are formulated under lossless waveguide assumptions. In this paper, we develop a unified physical model that jointly characterizes in-waveguide propagation loss and coupling effects. Specifically, we first establish an in-waveguide propagation model based on the EM field theory. Then, we obtain the corresponding waveguide-PA coupling model by using the coupled-mode theory.
    
The EM waves propagation in dielectric waveguides produces amplitude fading and phase changes. Both signal phase shift and propagation loss are related to the position of PA within the waveguide. The propagation constant of an EM waves through a dissipative dielectric material can be expressed as \cite{pozar2021microwave}:
\begin{equation}\label{eq:gamma}
{\gamma _g} = \sqrt {\kappa _c^2 - {\kappa ^2}} ,
\end{equation}
where $\kappa$ is the wave number of the EM waves in the waveguide, and $\kappa_c$ is the cut-off wave number satisfying $\kappa \ge \kappa_c$. It is worth noting that the propagation constant $\gamma_g$ is a complex number, which can be written as:
\begin{equation}\label{eq:complex number}
{\gamma _g} = {\alpha _g} + j{\beta _g} ,
\end{equation}
where ${\alpha _g}$ denotes the dielectric loss for EM waves propagation, and ${\beta _g}$ denotes the phase shift of the EM waves propagating in the waveguide. The dielectric loss and phase shift constant can be expressed as:
\begin{subequations}\label{eq:beta}
\begin{align}
{\alpha _g} &= \frac{{{\kappa^2}\tan \delta }}{{2{\beta _g}}}, \\
{\beta _g} &= \sqrt {{\kappa^2} - \kappa_c^2} ,
\end{align}
\end{subequations}
where $\tan \delta$ is the dielectric loss tangent. 

When condition $\kappa \gg {\kappa_c}$ is satisfied, the propagation loss and phase shift constants can be written as:
\begin{subequations}\label{eq:alpha}
\begin{align}
{\alpha _g} &= \frac{{\pi \sqrt {{\varepsilon _c}} \tan \delta }}{{{\lambda _0}}}  , \\
{\beta _g}  &= \frac{{2\pi \sqrt {{\varepsilon _c}} }}{{{\lambda _0}}} ,
\end{align}
\end{subequations}
where $\varepsilon _c$ is the relative permittivity, and $\lambda _0$ is the wavelength in free space. Thus, the signal propagating in the dielectric waveguide to position $x$ can be written as:
\begin{equation}\label{eq:sg}
{{\bf{s}}_g} = {e^{ - \left( {{\alpha_g}  + j{\beta _g} } \right){x}}}{s_0},
\end{equation}
where ${s_0}$ denotes the initial signal coding with $\mathbb{E}\left\{ {{{\left| {{s_0}} \right|}^2}} \right\} = 1$.

According to the in-waveguide propagation model \eqref{eq:sg}, the electric fields in waveguide and in PA can be expressed separately as follows \cite{pozar2021microwave}:
\begin{subequations}\label{eq:electric_field}
\begin{align}
&{\boldsymbol{{\mathop{\rm E}\nolimits} }_g}\left( {x,y,z} \right) = {e^{ - {\gamma _g}x}}{\boldsymbol{{\mathop{\rm D}\nolimits}}_g}\left( {y,z} \right){{\bf{s}}_0}, \\
&{\boldsymbol{{\mathop{\rm E}\nolimits}}_p}\left( {x,y,z} \right) = {e^{ - {\gamma _p}x}}{\boldsymbol{{\mathop{\rm D}\nolimits}}_p}\left( {y,z} \right){{\bf{s}}_0},
\end{align}
\end{subequations}
where $\boldsymbol{{\mathop{\rm D}\nolimits} _g}\left( {y,z} \right) \in {\mathbb{C}^{3 \times 1}}$ denotes the transverse field distribution in the waveguide, $\boldsymbol{{\mathop{\rm D}\nolimits} _p}\left( {y,z} \right) \in {\mathbb{C}^{3 \times 1}}$ denotes the transverse field distribution in PA, and ${\gamma _p}$ denotes the propagation coefficient in PA.

In this paper, the PA is modeled as an open-ended directional coupler, with a size of $L$. When the waveguide and PAs are placed in close proximity, a portion of the EM waves propagating within the waveguide is coupled into the PA \cite{okamoto2021fundamentals}. At this point, the size of PA is the coupling length.

According to the coupled-mode theory, the total EM field can be expressed as a weighted sum of the EM field in dielectric waveguide and the EM field in PA. Therefore, the total EM field can be expressed as:
\begin{equation}\label{eq:EM_field}
\boldsymbol{\mathop{\rm E}\nolimits}  = {{\mathop{\rm c}\nolimits} _g}\left( L \right)\boldsymbol{{\mathop{\rm E}\nolimits} _g} + {{\mathop{\rm c}\nolimits} _p}\left( L \right)\boldsymbol{{\mathop{\rm E}\nolimits} _p},
\end{equation}
where ${{\mathop{\rm c}\nolimits} _g}\left( L \right)$ and ${{\mathop{\rm c}\nolimits} _p}\left( L \right)$ denote the coupling coefficients, respectively.

By substituting the EM field expressions into the Maxwell’s equations, the following expressions for ${{{\mathop{\rm c}\nolimits} _g}\left( L \right)}$ and ${{{\mathop{\rm c}\nolimits} _p}\left( L \right)}$ can be derived as follows \cite{okamoto2021fundamentals}:
\begin{subequations}\label{eq:coupled_equations}
\begin{align}
&\frac{{d{{\mathop{\rm c}\nolimits} _g}\left( L \right)}}{{dL}} =  - j\chi {{\mathop{\rm c}\nolimits} _p}\left( L \right){e^{ - \Delta \gamma L}}, \\
&\frac{{d{{\mathop{\rm c}\nolimits} _p}\left( L \right)}}{{dL}} =  - j\chi {{\mathop{\rm c}\nolimits} _g}\left( L \right){e^{ - \Delta \gamma L}},
\end{align}
\end{subequations}
where $\chi$ is a mode coupling coefficient of the directional coupler, $\Delta \gamma  = {\gamma _p} - {\gamma _g}$ denotes the difference between the propagation constant of waveguide, and the propagation constant of PA. Since the EM waves propagates from the waveguide into the PA, the following initial conditions can be obtained:
\begin{subequations}\label{eq:condition}
\begin{align}
&{{\mathop{\rm c}\nolimits} _g}\left( 0 \right) = 1,\\
&{{\mathop{\rm c}\nolimits} _p}\left( 0 \right) = 0.
\end{align}
\end{subequations}

Solving \eqref{eq:coupled_equations} with initial conditions \eqref{eq:condition} gives the following expression:
\begin{subequations}\label{eq:AB}
\begin{align}
&{{\mathop{\rm c}\nolimits} _g}\left( L \right) = \left( {\cos \left( {\frac{{\phi L}}{2}} \right) + \frac{{\Delta \gamma }}{\phi }\sin \left( {\frac{{\phi L}}{2}} \right)} \right){e^{ - \frac{{\Delta \gamma L}}{2}}} , \\
&{{\mathop{\rm c}\nolimits} _p}\left( L \right) =  -j \frac{{2\chi }}{\phi }\sin \left( {\frac{{\phi L}}{2}} \right){e^{\frac{{\Delta \gamma L}}{2}}},
\end{align}
\end{subequations}
where $\phi  = \sqrt {4{\chi ^2}-\Delta {\gamma ^2}} $. 

In this paper, we consider a special case where the dielectric waveguide and PAs have identical propagation constants, i.e., ${\gamma _p} = {\gamma _g}$, which implies $\Delta \gamma  = 0$ and $\phi  = 2\chi $. The forms of ${{{\mathop{\rm c}\nolimits} _g}\left( L \right)}$ and ${{{\mathop{\rm c}\nolimits} _p}\left( L \right)}$ are simplified as follows:
\begin{subequations}\label{eq:delta0}
\begin{align}
&{{\mathop{\rm c}\nolimits} _g}\left( L \right) = \cos \left( {\chi L} \right), \\
&{{\mathop{\rm c}\nolimits} _p}\left( L \right) =  - j\sin \left( {\chi L} \right).
\end{align}
\end{subequations}

In summary, the EM field at the open end of PA can be written as:
\begin{subequations}\label{eq:ES}
\begin{align}
{\boldsymbol{{\mathop{\rm E}\nolimits}}_s}\left( {y,z} \right) &= {\boldsymbol{{\mathop{\rm E}\nolimits} }_g}\left( {r,y,z} \right){{\mathop{\rm c}\nolimits} _p}\left( L \right){e^{ - {\gamma _g} L }}\\
&= - j{\boldsymbol{{\mathop{\rm D}\nolimits}}_g}\left( {y,z} \right)\sin \left( {\chi {L}} \right){e^{ - {\gamma _g}\left( {{r} + {L}} \right)}}{s_0},
\end{align}
\end{subequations}
where ${r}$ denotes the position of left endpoint of PA, i.e., the propagation distance of the EM wave inside the waveguide before the coupling effect occurs. Once the coupling between the PA and waveguide ends, the residual EM field within waveguide is given by:
\begin{equation}\label{eq:remain}
\boldsymbol{{\mathop{\rm E}\nolimits}}_g^\prime \left( {x,y,z} \right) = {\boldsymbol{{\mathop{\rm D}\nolimits}}_g}\left( {y,z} \right)\cos \left( {\chi {L}} \right){e^{ - {\gamma _g}x }}{s_0}.
\end{equation}
\vspace{-6.9mm}
\subsection{Signal Model}

To avoid interference caused by overlapped signals, we exploit the orthogonality of time slots in time division multiple access (TDMA) to isolate interference. Specifically, we consider the case $K = TM$, where $\textit{T}$ is a positive integer. So each waveguide is paired with $T$ users, and in any given time slot each waveguide serves only one of its paired users. 

We introduce precoding to enable user scheduling and power allocation. The precoding vector of the $\textit{m}$-th waveguide at time slot $\textit{t}$ takes the form:
\begin{equation}\label{eq:mW_k}
{{\bf{w}}_m}\left( t \right) = {\left[ {{w_{1,m}}{\delta _{1,m}}{\delta _{1,t}}, \cdots ,{w_{K,m}}{\delta _{K,m}}{\delta _{K,t}}} \right]^T},
\end{equation}
where ${w_{k,m}}$ is the power allocation factor, ${\delta _{k,m}}$ denotes the pairing scheme between the waveguide and user, and ${\delta _{k,t}}$ denotes the user selection within time slot $\textit{t}$. It is worth noting that both ${\delta _{k,m}}$ and ${\delta _{k,t}}$ are binary zero-one variables. When ${\delta _{k,m}}{\delta _{k,t}} = 1$, it indicates that the $\textit{m}$-th waveguide serves the $\textit{k}$-th user in the $\textit{t}$-th time slot.

Therefore, the signals propagated in the $\textit{m}$-th waveguide at time slot $\textit{t}$ are as follows:
\begin{equation}\label{eq:mS_m}
{S_m}\left( t \right) = {{\bf{w}}_m}{\left( t \right)^T}{{\bf{C}}_k},
\end{equation}
where ${{\bf{C}}_k}$ is the signal coding of the $\textit{k}$ users. The precoding vector should obey the power constraint $\sum\limits_{m = 1}^M {{{\left\| {{{\bf{w}}_m\left( t \right)}} \right\|}^2}} \le P$, where $\textit{P}$ is the total transmit power budget for PASS. 

Owing to the unidirectional propagation of signals in the waveguide, the incident field at the $n$-th PA is the residual field remaining after the coupling of the preceding $n-1$ PAs. By recursively applying the field models \eqref{eq:ES} and \eqref{eq:remain}, the coupling coefficient of the $(m,n)$-th PA is derived as:
\begin{equation}\label{eq:xi_m,n}
\xi_{m,n} = 
\begin{cases} 
\sin(\chi L_{m,1}), & n=1, \\
\left(\displaystyle\prod_{i=1}^{n-1} \cos(\chi L_{m,i})\right) \sin(\chi L_{m,n}), & n \geq 2 .
\end{cases}
\end{equation}
Therefore, the signal radiated by the $(m,n)$-th PA in the $\textit{t}$-th time slot is formulated as:
\begin{subequations}\label{eq:sn}
\begin{align}
{s_{m,n}}\left( t \right) = {\xi _{m,n}}{e^{ - {\gamma _g}{x_{m,n}} - j\frac{\pi }{2}}}{S_m}\left( t \right).
\end{align}
\end{subequations}
Accordingly, the signal vector radiated by the $\textit{m}$-th waveguide in the $\textit{t}$-th time slot can be written as:
\begin{equation}\label{eq:s_m}
{{\bf{s}}_m}\left( t \right) = {\left[ {{s_{m,1}}\left( t \right), \cdots ,{s_{m,N}}\left( t \right)} \right]^T}.
\end{equation}

Based on the PA radiated signal model in \eqref{eq:sn}, we can change the characteristics of the radiation signal by controlling the mode coupling coefficient and coupling length. Therefore, in this paper, the PAs on each waveguide divide the signal power equally. The coupling coefficient is $\sin \left( {\chi {L_{m,n}}} \right) = \frac{1}{{\sqrt {N - n + 1} }}$. The signal radiated by the $\textit{n}$-th PA is given by:
\begin{equation}\label{eq:AP}
{s_{m,n}}\left( t \right) = \frac{1}{{\sqrt N }}{e^{ - {\gamma _g}{x_{m,n}} - j\frac{\pi }{2}}}{S_m}\left( t \right).
\end{equation}

\section{Joint Design and Optimization for PASS}

\subsection{Problem Formulation}

Assuming that the $\textit{k}$-th user is paired with the $\textit{m}$-th waveguide, the signal received by the $\textit{k}$-th user in the $\textit{t}$-th time slot can be expressed as:
\begin{equation}\label{eq:my_k}
{y_{m,k}}\left( t \right) = {\bf{h}}_{m,k}^T{{\bf{s}}_m}\left( t \right) + \underbrace {\sum\limits_{i = 1,i \ne m}^M {{\bf{h}}_{i,k}^T{{\bf{s}}_i}\left( t \right)} }_{{\rm{interference}}} + {n_k},
\end{equation}
where the interference arises from the PAs located on the remaining $M-1$ waveguides and ${n_k}$ is additive white Gaussian noise (AWGN) with noise power ${{\sigma ^2}}$. Accordingly, the information rate of the $\textit{k}$-th user can be written as:
\begin{equation}\label{eq:mR_k}
{R_k} = \frac{1}{T}{\log _2}\left( {1 + {{{\mathop{\rm SINR}\nolimits} }_{m,k}}\left( t \right)} \right),
\end{equation}
where ${{\mathop{\rm SINR}\nolimits} _{m,k}}\left( t \right) = \frac{{{{\left| {{\bf{h}}_{m,k}^T{{\bf{s}}_m}\left( t \right)} \right|}^2}}}{{{{\left| {\sum\limits_{i = 1,i \ne m}^M {{\bf{h}}_{i,k}^T{{\bf{s}}_i}\left( t \right)} } \right|}^2} + {\sigma ^2}}}$. 

Thus, the system sum rate of PASS can be expressed as follows:
\begin{equation}\label{eq:mR_sum}
{R_{sum}} = \sum\limits_{k = 1}^K {\frac{1}{T}{{\log }_2}\left( {1 + {{{\mathop{\rm SINR}\nolimits} }_{m,k}}\left( t \right)} \right)} .
\end{equation}

Since PAs are modeled as open directional couplers with a physical size of ${L_{m,n}}$, PAs' positions are discretely distributed along the waveguide rather than being concentrated at a single point. For tractability, PAs are positioned sequentially along the waveguide. The corresponding constraints are given by:
\begin{subequations}\label{eq:mposition}
\begin{align}
&{L_{m,1}} \le {x_{m,1}},{x_{m,N}} \le D, \\
&{x_{m,i}} - {x_{m,i - 1}} \ge {L_{m,i}},i = 2, \ldots ,N.
\end{align}
\end{subequations}
After simplification, the constraints of PAs' positions distribution are represented in a vector form as:
\begin{equation}\label{eq:mAX}
{{\bf{p}}_m} \ge {{\bf{a}}_m},
\end{equation}
where the vectors ${{\bf{p}}_m}$ and ${{\bf{a}}_m}$ are of the following form: 
\begin{subequations}\label{eq:A}
\begin{align}
{{\bf{p}}_m} &= {\left[ {{x_{m,1}},{x_{m,2}} - {x_{m,1}}, \cdots ,{x_{m,N}} - {x_{m,N - 1}}, - {x_{m,N}}} \right]^T}, \\
{{\bf{a}}_m} &= {\left[ {{L_{m,1}}, \cdots ,{L_{m,N}}, - D} \right]^T}.
\end{align}
\end{subequations}

The optimization problem for maximizing the sum rate of multi-waveguide PASS is given by:
\begin{subequations}\label{eq:msignal op}
\begin{align}
\mathop {\max }\limits_{{{\bf{w}}_m}, {\psi _{m,n}}}\;&{R_{sum}} , \\
s.t.\;&C1:{{\bf{p}}_m} \ge {{\bf{a}}_m},\forall m, \\
&C2:{R_k} \ge {R_{\min }}, \forall k,\\
&C3:\sum\limits_{m = 1}^M {{{\left\| {{{\bf{w}}_m}\left( t \right)} \right\|}^2}}  \le P,\\
&C4:{\delta _{k,m}},{\delta _{k,t}} = \left\{ {0,1} \right\},\forall k,m,t,\\
&C5a:\sum\limits_{m = 1}^M {{\delta _{k,m}}}  = 1,\forall k,\\
&C5b:\sum\limits_{k = 1}^K {{\delta _{k,m}}}  = T,\forall m,\\
&C6a:\sum\limits_{k = 1}^K {{\delta _{k,t}}}  = M,\forall t,\\
&C6b:\sum\limits_{t = 1}^T {{\delta _{k,t}}}  = 1,\forall k,
\end{align}
\end{subequations}
where ${R_{\min }}$ is the minimum rate. Constraint $C5a$ ensures that each user is paired with only one waveguide, while constraint $C5b$ specifies that each waveguide is paired with \textit{T} users. Constraint $C6a$ guarantees that each waveguide serves only one user per time slot, and constraint $C6b$ ensures that each user occupies only one time slot.
\vspace{-0.9mm}
\subsection{Hierarchical User Scheduling Algorithm}

The proposed HUS contains two layers: waveguide-user pairing and in-slot user selection. In the waveguide-user pairing layer, each user is associated with a geometrically suitable waveguide to reduce the large-scale fading. In the in-slot user selection layer, spatially separated users are preferentially scheduled in the same time slot, so that the interference signals are weakened by increasing the propagation distance. Therefore, the HUS algorithm jointly improves the sum rate from the aspects of path loss reduction and interference management.

\subsubsection{\textbf{Waveguide-user pairing algorithm}}Since PAs can only move along the waveguide, a smaller distance between the user and the waveguide generally leads to a lower path loss. Therefore, the distance between the user and the waveguide is adopted as the pairing metric, so that each user is paired with a geometrically suitable waveguide, thereby providing a better basis for the subsequent deployment of PAs. Accordingly, the pairing problem is formulated as:
\begin{subequations}\label{eq:mpair1}
\begin{align}
\mathop {\min }\limits_{{\delta _{k,m}}}\;&\sum\limits_{k = 1}^K {\sum\limits_{m = 1}^M {{\delta _{k,m}}d_{k,m}^2} }  , \\
s.t.\;&C4,C5a,C5b,
\end{align}
\end{subequations}
where $d_{k,m}$ is the geometric distance from the $\textit{k}$-th user to the $\textit{m}$-th waveguide.


In order to turn \eqref{eq:mpair1} into a convex optimization problem, the binary zero-one variable ${\delta _{k,m}}$ is relaxed to a continuous variable $C4':{{\delta '}_{k,m}} \in \left[ {0,1} \right]$. The corresponding constraints take the following forms:
\begin{subequations}\label{eq:c5a}
\begin{align}
&C5a':\sum\limits_{m = 1}^M {{{\delta '}_{k,m}}}  = 1,\forall k, \\
&C5b':\sum\limits_{k = 1}^K {{{\delta '}_{k,m}}}  = T,\forall m.
\end{align}
\end{subequations}
The problem is rewritten as:
\begin{subequations}\label{eq:mpair2}
\begin{align}
\mathop {\min }\limits_{{{\delta'} _{k,m}}}\;&\sum\limits_{k = 1}^K {\sum\limits_{m = 1}^M {{{\delta '}_{k,m}}d_{k,m}^2} }  , \\
s.t.\;&C4',C5a',C5b'.
\end{align}
\end{subequations}
The optimization problem \eqref{eq:mpair2} is convex and can be solved by using a convex programming solver. The actual pairing scheme can be obtained after integerization. The steps are shown in \textbf{Algorithm 1}.

The waveguide-user pairing problem is convex and solved optimally. Then, a feasible binary pairing scheme is obtained through the integerization. The complexity is dominated by solving a linear program with $KM$ variables, which requires  $O((KM)^3)$.

\begin{algorithm}[t]
\caption{Waveguide-user pairing algorithm}\label{algorithm1}
\begin{algorithmic}[1]
    \STATE The distance $d_{k,m}$ and $d_{i,k}$ is calculated based on the position of the users and dielectric waveguides;
    \STATE Solve problem \eqref{eq:mpair2} by CVX;
\STATE Initialize $\delta_{k,m} = 0$ for all $k,m$;
\REPEAT
    \FOR{each user $k$}
        \STATE Find the waveguide index with the largest $\delta'_{k,m}$, i.e., $m^* = \arg\max_{m} \delta'_{k,m}$; 
        \STATE Pair user $k$ with the $m^*$-th waveguide, i.e., $\delta_{k,m^*} = 1$;
    \ENDFOR
    \FOR{each waveguide with $\sum_k {{\delta _{k,m}}} > T$}
        \STATE Sort users paired with the $m$-th waveguide by $\delta'_{k,m}$;
        \STATE Remove the users  with smallest $\delta'_{k,m}$ by setting the corresponding $\delta_{k,m} = 0$;
        \STATE Reset the corresponding $\delta'_{k,m} = 0$ for removed users;
    \ENDFOR
\UNTIL{$\sum_m {{\delta _{k,m}}}  = 1, \forall k$ and $\sum_k {{\delta _{k,m}}} = T, \forall m$}
\end{algorithmic}
\end{algorithm}

\subsubsection{\textbf{In-slot user selection algorithm}}
After the pairing stage, each dielectric waveguide serves its paired users based on the TDMA architecture. However, the signals radiated by PAs on different waveguides in the same time slot introduce mutual interference. Therefore, the in-slot user selection has to take the interference relationship among candidate users into account. In addition, to reduce the impact of large-scale fading, the PAs in PASS are usually deployed close to the served users. Based on the proximity deployment, spatially separated users are preferentially scheduled in the same time slot, since such a selection increases the propagation distance of interference signals, thereby reducing inter-waveguide interference and improving the transmission efficiency.

Based on the above observations, the in-slot user selection problem is formulated in a SINR-inspired form, so that the selected users can achieve a more favorable balance between useful signal strength and inter-waveguide interference. Accordingly, the in-slot user selection problem is formulated as:
\begin{subequations}\label{eq:mselection1}
\begin{align}
\mathop {\max }\limits_{{\delta _{k,t}}}\;&\sum\limits_{k = 1}^K {\frac{1}{T}{{\log }_2}\left( {1 + \frac{1}{{d_{k,m}^2\left({\sigma ^2}+ {\sum\limits_{i = 1,i \ne k}^K {\frac{{{\delta _{i,t}}}}{{d_{i,k}^2}}} } \right)}}} \right)} , \\
s.t.\;&C4,C6a,C6b,
\end{align}
\end{subequations}
where $d_{i,k}$ represents the distance between the $i$-th user and the $k$-th user.

For the sake of simplicity, the objective function is expressed as:
\begin{equation}\label{eq:mfkt}
\begin{split}
{f_{k,t}} = \frac{1}{T}\left( {{{\log }_2}\left( {{I_{k,t}} + 1} \right) - {{\log }_2}\left( {{I_{k,t}}} \right)} \right),
\end{split}
\end{equation}
where ${I_{k,t}} = d_{k,m}^2\left( {{\sigma ^2} + \sum\limits_{i = 1,i \ne k}^K {\frac{{{\delta _{i,t}}}}{{d_{i,k}^2}}} } \right)$. 

To address the non-concavity of objective function, the first-order Taylor expansion of ${ - {{\log }_2}\left( {I_{k,t}} \right)}$ can be expressed as:
\begin{equation}\label{eq:mfktsca}
\begin{split}
{f_{s}}\left( { - {{\log }_2}\left( {{I_{k,t}}} \right)} \right)  \buildrel \Delta \over =   &- {\log _2}\left( {I_{k,t}^{\left( {{t_1}} \right)}} \right)\\
 -& \sum\limits_{i \ne k} {\frac{{d_{k,m}^2}}{{\left(\ln 2\right)I_{k,t}^{\left( {{t_1}} \right)}d_{i,k}^2}}\left( {{\delta _{i,t}} - \delta _{i,t}^{\left( {{t_1}} \right)}} \right)} ,
\end{split}
\end{equation}
where $t_1$ is the iteration index. At the same time, the non-convex optimization variable is relaxed to a continuous variable $\overline{C4}:{{\delta '}_{k,t}} \in \left[ {0,1} \right]$. The corresponding constraints can be expressed as:
\begin{subequations}\label{eq:c6a}
\begin{align}
&C6a':\sum\limits_{k = 1}^K {{{\delta'} _{k,t}}}  = M,\forall t, \\
&C6b':\sum\limits_{t = 1}^T {{{\delta'} _{k,t}}}  = 1,\forall k.
\end{align}
\end{subequations}

Now, the in-slot user selection problem is rewritten as:
\begin{subequations}\label{eq:mselection2}
\begin{align}
\mathop {\max }\limits_{{{\delta'} _{k,t}}}\;&\sum\limits_{k = 1}^K {\frac{1}{T}\left( {{{\log }_2}\left( {{{I'}_{k,t}} + 1} \right) + {f_s}\left( { - {{\log }_2}\left( {{{I'}_{k,t}}} \right)} \right)} \right)} , \\
s.t.\;&\overline{C4},C6a',C6b'.
\end{align}
\end{subequations}
Problem \eqref{eq:mselection2} can be solved iteratively by using a standard convex programming solver. The optimal user selection scheme can be obtained after integerization. The user selection algorithm step is shown in \textbf{Algorithm 2}.

Since ${I_{k,t}}$ is an affine function of the optimization variables, $- {{\log }_2}\left( {{I_{k,t}}}\right)$ is convex. Therefore, the first-order Taylor expansion in \eqref{eq:mfktsca} gives a global lower bound for the original function, and the objective in \eqref{eq:mselection2} serves as a lower-bound surrogate of the original objective. Moreover, since the approximated term is obtained by first-order Taylor expansion, the surrogate is locally tight at the current iterate, meaning that it shares the same function value and first-order derivative as the original objective. Hence, the objective value is monotonically non-decreasing over the iterations, and the proposed algorithm converges to a stationary point. The complexity of the algorithm is $O(I_1(KT)^3)$, since each SCA iteration solves a convex program with $KT$ variables, where $I_1$ is the number of SCA iterations.

\begin{algorithm}[t]
\caption{In-slot user selection algorithm}\label{algorithm2}
\begin{algorithmic}[1]
    \STATE Set the upper iteration limit $t_{1\max}$, and initialize the user selection variable $\delta _{k,t}^{\left( {{t_1}} \right)}$ and iteration index ${t_1}=0$;
    \WHILE {No convergence or ${t_1} \le t_{1\max}$}
        \STATE Solving problem \eqref{eq:mselection2} with CVX;
        \STATE Set ${t_1}={t_1}+1$;
        \STATE Update $\delta _{k,t}^{\left( {{t_1}} \right)}={{\delta'} _{k,t}}$;
    \ENDWHILE
    \STATE Initialize $\delta_{k,t} = 0$ for all $k,t$;
\REPEAT
    \FOR{each user $k$}
        \STATE Find the index with the largest $\delta_{k,t}^{(t_1)}$, i.e., $t^* = \arg\max_{t} \delta_{k,t}^{(t_1)}$;
        \STATE Set $\delta_{k,t^*} = 1$;
    \ENDFOR
    \FOR{each time slot $t$ with $\sum_k {{\delta _{k,t}}} > M$}
        \STATE Sort users by $\delta_{k,t}^{(t_1)}$;
        \STATE Remove users with smallest $\delta_{k,t}^{(t_1)}$ by setting the corresponding $\delta_{k,t} = 0$;
        \STATE Reset the corresponding $\delta_{k,t}^{(t_1)} = 0$ for removed users;
    \ENDFOR
\UNTIL{$\sum_t {{\delta _{k,t}}} = 1, \forall k$ and $\sum_k {{\delta _{k,t}}} = M, \forall t$}
\end{algorithmic}
\end{algorithm}
\vspace{-0.9mm}
\subsection{Alternating Optimization Algorithm} 
For a given user scheduling, maximizing the system sum rate is achieved by optimizing the positions of PAs and the power allocation factor. The optimization problem can be expressed as:
\begin{subequations}\label{eq:msignal op2}
\begin{align}
\mathop {\max }\limits_{{{w_{k,m}}},{\psi _{m,n}}}\;&{R_{sum}} , \\
s.t.\;&C1,C2,C3.
\end{align}
\end{subequations}
Based on the AO algorithm, optimization problem \eqref{eq:msignal op2} is decoupled into two sub-problems : optimization of PAs' positions and power allocation.

\subsubsection{\textbf{Optimization of the positions of PAs}}
For a given pairing scheme and power allocation, the optimization subproblem for the positions of PAs can be written as:
\begin{subequations}\label{eq:mxop1}
\begin{align}
\mathop {\max }\limits_{{\psi _{m,n}}}\;&\sum\limits_{k = 1}^K {{R_k}}, \\
s.t.\;&C1,C2.
\end{align}
\end{subequations}

All optimization variables in problem \eqref{eq:mxop1} are mutually coupled, making it difficult to obtain a global optimal solution directly. Therefore, based on the Gauss-Seidel algorithm \cite{bereyhi2025mimopassuplinkdownlinktransmission}, each position is optimized individually to achieve a near-optimal solution. The subproblem for the $(m,n)$-th PA's position takes the following form:
\begin{subequations}\label{eq:mxop2}
\begin{align}
\mathop {\max }\limits_{{x_{m,n}}}\;&\sum\limits_{k = 1}^K {\frac{1}{T}{{\log }_2}\left( {1 + {{\mathop{\rm SINR}\nolimits} _{m,k}\left( t \right)}} \right)} , \\
s.t.\;&C1,C2.
\end{align}
\end{subequations}

The optimized position $x_{m,n}$ can be achieved by using one-dimensional grid search. The deployable interval on the $\textit{m}$-th waveguide is discretised into $\textit{G}$ sampling intervals, with the midpoint of each interval representing the interval coordinates. Then, the collection of candidate positions on the $\textit{m}$-th waveguide can be written as:
\begin{equation}\label{eq:mG1}
\begin{split}
\mathbb{L}_m \buildrel \Delta \over = \left\{ {\frac{D}{{2G}},\frac{{3D}}{{2G}}, \cdots ,\frac{{\left( {2G - 1} \right)D}}{G}} \right\}.
\end{split}
\end{equation}

Considering the PAs' positions constraint $C1$, the feasible deployment interval of the $(m,n)$-th PA is jointly determined by the boundaries of its adjacent PAs to ensure their sequential arrangement. Therefore, the feasible deployment interval of the $(m,n)$-th PA is given by $\left[ {{x_{m,n - 1}}+ {L_{m,n}},{x_{m,n + 1}} - {L_{m,n + 1}}} \right]$. Only the candidate positions whose midpoint locations lie within the feasible deployment interval are regarded as feasible positions. The indices of all feasible positions are collected in $\mathbb{I}_{m,n}$. 

Then, the feasible positions for the one-dimensional search of the $(m,n)$-th PA can be expressed as:
\begin{equation}\label{eq:mG2}
\begin{split}
\mathbb{L}_{m,i} = \frac{{\left( {2i - 1} \right)D}}{{2G}},i \in \mathbb{I}_{m,n},
\end{split}
\end{equation}
The local optimal position can be obtained via a one-dimensional search over all feasible positions. Specifically, the achievable sum rate at each feasible position are evaluated, and the position that yields the maximum sum rate while satisfying constraint $C2$ is selected as the updated PA position. The detailed PAs' positions optimization procedure is summarized in \textbf{Algorithm 3}.

Since each step selects the best point from a finite discrete candidate set, the objective value is monotonically non-decreasing after each update. Furthermore, since the discretized feasible set is finite, the proposed one-dimensional search procedure is guaranteed to converge to a locally optimal discrete solution. The computational complexity is $O(MNG)$.

It is worth noting that the main suboptimality of the proposed AO framework comes from the PAs' positions optimization step, where the continuous feasible interval is discretized into $G$ candidate points for one-dimensional search. Therefore, only a near-optimal discrete solution can be obtained. As $G$ increases, the discretization induced gap decreases, at the cost of higher computational complexity.

\begin{algorithm}[t]
\caption{Optimization algorithm for PAs' positions}\label{algorithm3}
\begin{algorithmic}[1]
    \STATE Given the current power allocation coefficients $\{w_{k,m}\}$ and PA positions $\{\psi_{m,n}\}$;
    \FORALL{$m \in \{1, \ldots, M\}$}
        \FORALL{$n \in \{1, \ldots, N\}$}
            \STATE Based on the given $\{x_{m,n}\}$, the set of feasible interval numbers $\mathbb{I}_{m,n}$ is obtained;
            \STATE Update $x_{m,n}$ via one-dimensional search over problem \eqref{eq:mxop2};
        \ENDFOR
    \ENDFOR
    \STATE The updated PA positions are obtained as $\{\psi_{m,n}\}$;
\end{algorithmic}
\end{algorithm}

\subsubsection{\textbf{Optimization of power allocation}}
For a given set of PAs' positions, the optimization problem for power allocation consists of the following:
\begin{subequations}\label{eq:mwop1}
\begin{align}
\mathop {\max }\limits_{{{\bf{w}}_m}}\;&\sum\limits_{k = 1}^K {{R_k}} , \\
s.t.\;&C2,C3.
\end{align}
\end{subequations}

Since each waveguide serves only one user within a given time slot, the power allocation factors for each waveguide may vary across different time slots. The SINR of the signal transmitted by the $\textit{m}$-th waveguide to the $\textit{k}$-th user at time slot $\textit{t}$ can be rewritten as:
\begin{equation}\label{eq:msinr}
{{\mathop{\rm SINR}\nolimits} _{m,k}} = \frac{{{{\left| {{\bf{h}}_{m,k}^T{{\bf{g}}_m}{w_{k,m}}} \right|}^2}}}{{{{\left| {\sum\limits_{i = 1,i \ne m}^M {{\bf{h}}_{i,k}^T{{\bf{g}}_i}{w_{k,i}}} } \right|}^2} + {\sigma ^2}}},
\end{equation}
where ${{\bf{g}}_m} = {\left[ {{g_{m,1}}, \ldots ,{g_{m,N}}} \right]^T}$ is the vector of waveguide propagation loss and coupling effects, i.e., ${g_{m,n}} = {\xi _{m,n}}{e^{ - {\gamma _g}{x_{m,n}} - j\frac{\pi }{2}}}$. 

To address the non-convexity of optimization problem, we employ the Lagrange duality transformation \cite{8310563} and FP \cite{8314727} to introduce auxiliary variables. Then, we propose an iterative optimization algorithm based on SCA to obtain the optimal power allocation scheme.
%

\begin{theorem}\label{theorem1}
The weighted logarithmic summation function can be deformed as:
\begin{equation}\label{eq:mRsum1}
\begin{split}
{R_{sum}} = &\sum\limits_{k = 1}^K {\frac{1}{T}{{\log }_2}\left( {1 + {\nu _k}} \right)}  - \sum\limits_{k = 1}^K {\frac{{{\nu _k}}}{{T\ln 2}}}   \\
&+ \underbrace {\sum\limits_{k = 1}^K {\frac{{\left( {1 + {\nu _k}} \right){a_{m,k}}}}{{T\ln 2\left( {{a_{m,k}} + {b_{m,k}}} \right)}}} }_{Sum ~ of ~ ratio~{\rm{ }}term},
\end{split}
\end{equation}
where ${{\mathop{\rm a}\nolimits} _{m,k}}$ is the signal power, ${{\mathop{\rm b}\nolimits} _{m,k}}$ is the power of interference plus noise, and $\nu _k$ is an auxiliary variable updated with iteration. The auxiliary variables can be expressed as:
\begin{subequations}\label{eq:mrk}
\begin{align}
&{{\mathop{\rm a}\nolimits} _{m,k}} = {\left| {{\bf{h}}_{m,k}^T{{\bf{g}}_m}{w_{k,m}}} \right|^2}, \\
&{{\mathop{\rm b}\nolimits} _{m,k}} = {\left| {\sum\limits_{i = 1,i \ne m}^M {{\bf{h}}_{i,k}^T{{\bf{g}}_i}{w_{k,i}}} } \right|^2} + {\sigma ^2},\\
&{\nu _k} = \frac{{{{\mathop{\rm a}\nolimits} _{m,k}}}}{{{{\mathop{\rm b}\nolimits} _{m,k}}}}.
\end{align}
\end{subequations}
\begin{proof}
Please refer to Appendix A.
\end{proof}
\end{theorem}

\begin{remark}
By applying \textbf{Theorem 1}, the sum rate expression in logarithmic form \eqref{eq:mR_sum} is transformed into an equivalent fractional form \eqref{eq:mRsum1}, which is more tractable for convex optimization. Moreover, the information rate $\frac{1}{T}{\log _2}\left( {1 + {{{\mathop{\rm SINR}\nolimits} }_{m,k}}\left( t \right)} \right)$ can be obtained when the variable ${\nu _k}$ is set to its optimal value $\frac{{{{\mathop{\rm a}\nolimits} _{m,k}}}}{{{{\mathop{\rm b}\nolimits} _{m,k}}}}$.
\end{remark}

\begin{theorem}\label{theorem2}
The function associated with the optimization variable can be expressed as:
\begin{equation}\label{eq:mrk2}
\begin{split}
\frac{{{{\mathop{\rm a}\nolimits} _{m,k}}}}{{{{\mathop{\rm a}\nolimits} _{m,k}} + {{\mathop{\rm b}\nolimits} _{m,k}}}} = 2{y_k}\sqrt {{{\mathop{\rm a}\nolimits} _{m,k}}}  - y_k^2\left( {{{\mathop{\rm a}\nolimits} _{m,k}} + {{\mathop{\rm b}\nolimits} _{m,k}}} \right),
\end{split}
\end{equation}
where
\begin{equation}\label{eq:myk}
{y_k} = \frac{{\sqrt {{{\mathop{\rm a}\nolimits} _{m,k}}} }}{{{{\mathop{\rm a}\nolimits} _{m,k}} + {{\mathop{\rm b}\nolimits} _{m,k}}}}.
\end{equation}
\begin{proof}
Please refer to Appendix B.
\end{proof}
\end{theorem}

\begin{remark}
The fractional objective function is further reformulated into a quadratic form \eqref{eq:mrk2}. The equivalent quadratic representation eliminates the coupling in denominator, making the problem more amenable to iterative optimization.
\end{remark}


Based on \textbf{Theorem 1} and \textbf{Theorem 2}, the optimization problem for power allocation \eqref{eq:mwop1} can be rewritten as:
\begin{subequations}\label{eq:mwop2}
\begin{align}
\mathop {\max }\limits_{{{w_{k,m}}}}\;\sum\limits_{k = 1}^K &{\frac{{\left( {1 + {\nu _k}} \right)}}{T\ln 2}\left( {2{y_k}\sqrt {{{\mathop{\rm a}\nolimits} _{m,k}}}  - y_k^2\left( {{{\mathop{\rm a}\nolimits} _{m,k}} + {{\mathop{\rm b}\nolimits} _{m,k}}} \right)} \right)} , \\
s.t.\;C3:&\sum\limits_{m = 1}^M {{{\left\| {{{\bf{w}}_m}\left( t \right)} \right\|}^2}} \le  P,\\
C2':&{{\mathop{\rm a}\nolimits} _{m,k}} \ge {{\mathop{\rm b}\nolimits} _{m,k}}\left( {{2^{T{R_{\min }}}} - 1} \right),\forall k,m.
\end{align}
\end{subequations}

To address the non-concavity of the objective function, we apply the first-order Taylor expansion to obtain the approximate function as follows:
\begin{equation}\label{eq:mSCA}
\begin{split}
 {f_{s}}\left( {\sqrt {{{\mathop{\rm a}\nolimits} _{m,k}}} } \right) &\buildrel \Delta \over = \sqrt {{\mathop{\rm a}\nolimits} _{m,k}^{\left( t _2\right)}}  \\
+ &{\mathop{\rm Re}\nolimits} \left\{ {\frac{{{{\left|{{\bf{h}}_{m,k}^T{{\bf{g}}_m}}\right|}^2}}}{{2\sqrt {{\mathop{\rm a}\nolimits} _{m,k}^{\left( t _2\right)}} }}{{\left( {w_{k,m}^{\left( t _2\right)}} \right)}^*}\left( {{w_{k,m}} - w_{k,m}^{\left( t _2\right)}} \right)} \right\},
\end{split}
\end{equation}
where $t_2$ is the iteration index. The objective function can be expressed as:
\begin{equation}\label{eq:mRsca}
\begin{split}
{f_s}\left( {{w_{k,m}}} \right) = 2{y_k}{f_s}\left( {\sqrt {{{\mathop{\rm a}\nolimits} _{m,k}}} } \right) - y_k^2\left( {{{\mathop{\rm a}\nolimits} _{m,k}} + {{\mathop{\rm b}\nolimits} _{m,k}}} \right).
\end{split}
\end{equation}
Since function ${\sqrt {{{\mathop{\rm a}\nolimits} _{m,k}}} }$ can be expressed in the following weighted quadratic-norm form as follows:
\begin{equation}\label{eq:sqrta}
\begin{split}
\sqrt {{{\mathop{\rm a}\nolimits} _{m,k}}}  = \left| {{\bf{h}}_{m,k}^T{{\bf{g}}_m}} \right|\left| {{w_{k,m}}} \right|,
\end{split}
\end{equation}
it is a convex function with respect to the optimization variable. Therefore, the first-order Taylor expansion in \eqref{eq:mRsca} serves as a tight lower-bound surrogate of function ${\sqrt {{{\mathop{\rm a}\nolimits} _{m,k}}} }$. Moreover, since ${{\mathop{\rm a}\nolimits} _{m,k}}$ and ${{\mathop{\rm b}\nolimits} _{m,k}}$ are convex functions, the resulting objective function constitutes a lower-bound surrogate of the original objective function.

To address the non-convexity of constraint $C2'$, we apply the first-order Taylor expansion to the function ${{\mathop{\rm a}\nolimits} _{m,k}}$, which is given by:
\begin{equation}\label{eq:mRsca2}
\begin{split}
{f_{s}}\left( {{{\mathop{\rm a}\nolimits} _{m,k}}} \right)  &\buildrel \Delta \over =   {\mathop{\rm a}\nolimits} _{m,k}^{\left( t _2\right)} \\
+ &{\mathop{\rm Re}\nolimits} \left\{ {{{\left| {{\bf{h}}_{m,k}^T{{\bf{g}}_m}} \right|}^2}{{\left( {w_{k,m}^{\left( t _2\right)}} \right)}^*}\left( {{w_{k,m}} - w_{k,m}^{\left( t _2\right)}} \right)} \right\}.
\end{split}
\end{equation}
Since function ${{\mathop{\rm a}\nolimits} _{m,k}}$ is convex, its expansion in \eqref{eq:mRsca2} provides a tight lower bound. Accordingly, the constraint can be expressed as:
\begin{equation}\label{eq:mC22}
\begin{split}
{{\mathop{\rm a}\nolimits} _{m,k}} \ge {f_s}\left( {{{\mathop{\rm a}\nolimits} _{m,k}}} \right) \ge {{\mathop{\rm b}\nolimits} _{m,k}}\left( {{2^{T{R_{\min }}}} - 1} \right).
\end{split}
\end{equation}
Considering that function ${{\mathop{\rm b}\nolimits} _{m,k}}$ is convex and the coefficient ${{2^{z{R_{\min }}}} - 1}$ is nonnegative, constraint $C2'$ can be transformed into the following convex constraint $\overline{C2}'$:
\begin{equation}\label{eq:mC222}
\begin{split}
\overline{C2}': {{\mathop{\rm b}\nolimits} _{m,k}}\left( {{2^{z{R_{\min }}}} - 1} \right) - {f_s}\left( {{{\mathop{\rm a}\nolimits} _{m,k}}} \right) \le 0.
\end{split}
\end{equation}

Now, the optimization problem is given by:
\begin{subequations}\label{eq:mwop3}
\begin{align}
\mathop {\max }\limits_{{{w_{k,m}}}}\;&\sum\limits_{k = 1}^K {\frac{{\left( {1 + {\nu _k}} \right)}}{T\ln 2}{f_s}\left( {{w_{k,m}}} \right)} , \\
s.t.\;&\overline{C2}',C3.
\end{align}
\end{subequations}

In summary, the surrogate objective function in \eqref{eq:mwop3} is derived from the first-order Taylor expansions in \eqref{eq:mRsca} and constitutes a tight lower bound of the original non-convex objective. Since function ${{\mathop{\rm a}\nolimits} _{m,k}}$, and ${{\mathop{\rm b}\nolimits} _{m,k}}$ are all convex functions with respect to the optimization variables, the surrogate objective is concave and therefore the maximization in \eqref{eq:mwop3} is a convex optimization problem. Moreover, constraint $\overline{C2}'$ is convex as established in \eqref{eq:mC222}, and constraint $C3$ is a convex constraint. Consequently, problem \eqref{eq:mwop3} can be efficiently solved by using a standard convex programming solver.

It is worth noting that the auxiliary variables ${\nu _k}$ and ${y _k}$ are fixed values in iteration, determined by ${\psi _{m,n}}$ and $w_{k,m}$ given in the previous round. Therefore, the auxiliary variables can be expressed as:
\begin{subequations}\label{eq:mgammay}
\begin{align}
\nu _k^{\left( {t_2}+1 \right)} &= \frac{{a_{m,k}^{\left( {t_2 } \right)}}}{{b_{m,k}^{\left( {t_2 } \right)}}},\\
{{y_k}}^{\left(  {t_2}+1 \right)} &= \frac{{\sqrt {a_{m,k}^{\left( {t _2} \right)}} }}{{a_{m,k}^{\left( {t_2 } \right)} + b_{m,k}^{\left( {t _2} \right)}}}.
\end{align}
\end{subequations}
The power allocation algorithm is shown in \textbf{Algorithm 4}.

\begin{algorithm}[t]
\caption{Power allocation optimization algorithm}\label{algorithm4}
\begin{algorithmic}[1]
    \STATE Set ${t_2} = 0$, set the iteration threshold $t_{2\max}$;
    \STATE Set the current power allocation coefficients as $w_{k,m}^{\left( t _2\right)}$;
    \WHILE {No convergence or ${t_2} \le t_{2\max}$}
        \STATE Update intermediate variables ${\mathop{\rm a}\nolimits} _{m,k}^{\left( t _2\right)}$ and ${\mathop{\rm b}\nolimits} _{m,k}^{\left( t _2\right)}$;
        \STATE Updating auxiliary variables $\nu _k^{\left( {t_2}+1 \right)}$ and ${\mathop{\rm y}\nolimits} _k^{\left( {t_2}+1 \right)}$ by \eqref{eq:mgammay};
        \STATE Solving optimization problem \eqref{eq:mwop3} via CVX;
        \STATE Updating optimization variable $w_{k,m}^{\left( {t_2}+1 \right)}$;
        \STATE Set $ {t_2}= {t_2}+1$;
    \ENDWHILE
    \STATE The updated power allocation coefficients are obtained as $w_{k,m}^{\left( t _2\right)}$;
\end{algorithmic}
\end{algorithm}

At each iteration, the non-convex objective and constraints are replaced with first-order Taylor expansions. Specifically, the surrogate objective serves as a global lower bound of the original objective, while the surrogate constraints form inner approximations of the original feasible set. Moreover, these approximations are locally tight at the current iterate, i.e., they coincide with the original functions in both function value and first-order derivative at the current operating point. As a result, the current iterate remains feasible to the surrogate problem, and solving the surrogate problem optimally yields a non-decreasing objective value over iterations. Since the objective value is also upper bounded by the finite transmit power budget, the proposed iterative procedure is guaranteed to converge. Under standard SCA conditions, any limit point of the generated sequence satisfies the Karush-Kuhn-Tucker (KKT) conditions of the original problem.

For the power-allocation step, the approximation is introduced through locally tight first-order surrogate functions. Therefore, the resulting suboptimality mainly stems from the convergence of the iterative procedure to a KKT point of the original non-convex problem, rather than from a significant approximation error introduced by the surrogate construction itself. The computational complexity is dominated by solving a convex program with $MT$ variables in each SCA iteration, yielding an overall complexity of $O(I_2(MT)^3)$, where $I_2$ is the number of iterations.

\subsection{Overall Algorithmic Framework}

\begin{algorithm}[t]
\caption{Overall Algorithmic Framework for Multi-Waveguide PASS}
\label{alg:overall_framework}
\begin{algorithmic}[1]
\REQUIRE User locations $\{\mathbf{U}_k\}_{k=1}^K$, waveguide parameters, total transmit power $P$, minimum rate requirement $R_{\min}$, maximum iteration number $I_{\max}$, tolerance $\epsilon$
\ENSURE User pairing variables $\delta_{k,m}$, in-slot scheduling variables $\delta_{k,t}$, PA positions $\{\psi_{m,n}\}$, and power allocation coefficients $\{w_{k,m}\}$
\STATE Initialize the PA positions $\{\psi_{m,n}^{(0)}\}$ and power allocation coefficients $\{w_{k,m}^{\left( 0 \right)}\}$;
\STATE Obtain the waveguide-user pairing variables $\delta_{k,m}$ by Algorithm~\ref{algorithm1};
\STATE Obtain the in-slot user selection variables $\delta_{k,t}$ by Algorithm~\ref{algorithm2};
\STATE Set the outer iteration index $i=0$;
\STATE Compute the initial sum rate $R_{\mathrm{sum}}^{(0)}$;
\REPEAT
    \STATE Given $\{w_{k,m}^{\left( i \right)}\}$, update the PA positions $\{\psi_{m,n}^{(i+1)}\}$ by Algorithm~\ref{algorithm3};
    \STATE Given $\{\psi_{m,n}^{(i+1)}\}$, update the power allocation coefficients $\{w_{k,m}^{\left( i+1 \right)}\}$ by Algorithm~\ref{algorithm4};
    \STATE Compute the updated sum rate $R_{\mathrm{sum}}^{(i+1)}$;
    \STATE Set $i \leftarrow i+1$;
\UNTIL{$|R_{\mathrm{sum}}^{(i)}-R_{\mathrm{sum}}^{(i-1)}| \le \epsilon$ or $i \ge I_{\max}$}
\STATE \textbf{return} $\delta_{k,m}$, $\delta_{k,t}$, $\{\psi_{m,n}\}$, and $\{w_{k,m}\}$
\end{algorithmic}
\end{algorithm}

Based on the proposed HUS algorithm and AO-based joint optimization strategy, the overall algorithmic framework for solving problem \eqref{eq:msignal op} is summarized in \textbf{Algorithm 5}. Since the considered problem is non-convex, a conventional initialization strategy is adopted in the proposed framework. Specifically, the PAs' positions are initially set to be uniformly distributed over the middle section of each waveguide, i.e., within the interval $\left[ {\frac{D}{4},\frac{{3D}}{4}} \right]$, while the power allocation coefficients are initialized as $\sqrt {\frac{P}{M}} $, corresponding to an equal transmit power allocation across all waveguides. The proposed framework efficiently coordinates user scheduling and joint optimization, thereby improving the system sum rate while satisfying the users' minimum rate requirements.

As analyzed above, the PAs' positions optimization step yields a non-decreasing objective value through one-dimensional search, while the power-allocation step also guarantees monotonic improvement because the convex surrogate is locally tight. Therefore, when the PAs' positions optimization and the power allocation are alternately executed, the overall objective value remains non-decreasing throughout the iterations. Moreover, under the finite transmit power budget, the achievable sum rate is upper bounded. Hence, the proposed overall algorithm is guaranteed to converge.

Moreover, the optimality gap of the proposed algorithm mainly stems from the discretized PAs' positions search and the local nature of the SCA-based power allocation update. Nevertheless, with a sufficiently fine search resolution and locally tight surrogate construction, the optimality gap is expected to remain limited in practice.

\section{Numerical Results}

In this section, we perform numerical simulations to evaluate the impact of in-waveguide propagation loss and coupling effects, and to verify the effectiveness of the proposed algorithm. The dielectric waveguide and PAs are assumed to be made of polytetrafluoroethylene, whose the key parameters are ${\varepsilon _c} = 2.08$ and $\tan \delta  = 0.0004$ \cite{pozar2021microwave}. The communication bandwidth is set to 1 MHz. According to the noise power spectral density ${P_n} =  - 174 + 10{\log _{10}}\left( B \right)$ dBm at room temperature, the AWGN power ${{\sigma ^2}}$ is set to $-114$ dBm. 

Two benchmark schemes are considered in this section: 1) Ideal waveguide scheme (IWS): IWS neglects both the propagation loss in the dielectric waveguide and the coupling effect between the waveguide and the PAs, thereby serving as an idealized upper bound for sum rate; 2) Dissipative waveguide scheme (DWS): DWS considers channel fading and in-waveguide propagation loss, which can evaluate the characterization of in-waveguide propagation loss. Notably, both IWS and DWS allow for the deployment of multiple PAs at the same position. The actual waveguide scheme (AWS) corresponds to the system model in this paper, incorporating both in-waveguide propagation loss and coupling effects. In particular, the comparison between IWS and DWS quantifies the effect of in-waveguide propagation loss, while the comparison between DWS and AWS further isolates the additional effect of coupling effect.

Unless otherwise specified, the following numerical results are obtained under a default multi-waveguide scenario, where three waveguides serve 9 users with PAs deployed on each waveguide, i.e., $K=9$, $M=3$, and $N=5$. The waveguide-related parameters are set as $D=10$ m, $h=3$ m, and $D_{\mathrm{wg}}=10$ m. The user distribution region is specified by $d_1=10$ m and $d_2=30$ m. For simulation, the service area is divided into nine subregions, each containing one randomly generated user. The minimum rate requirement of each user is set to $0.5$ bps/Hz. Unless otherwise stated, the reported sum rate is averaged over 100 Monte Carlo realizations. In addition, the resolution of the one-dimensional search is set to $G=10000$.
\vspace{-3.9mm}
\subsection{Convergence Behavior and Initialization Robustness of the Proposed Algorithm}

\begin{figure}[t]
    \centering
    \includegraphics[width=3.0in]{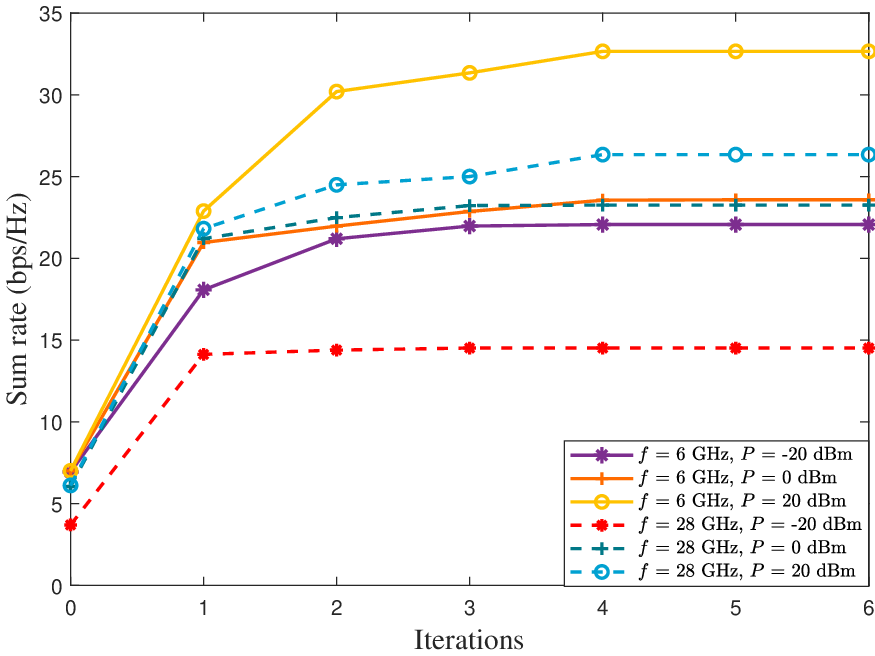} 
    \caption{Sum rate versus number of iterations.} 
    \label{figure4} 
\end{figure}

\begin{figure}[t]
    \centering
    \includegraphics[width=3.0in]{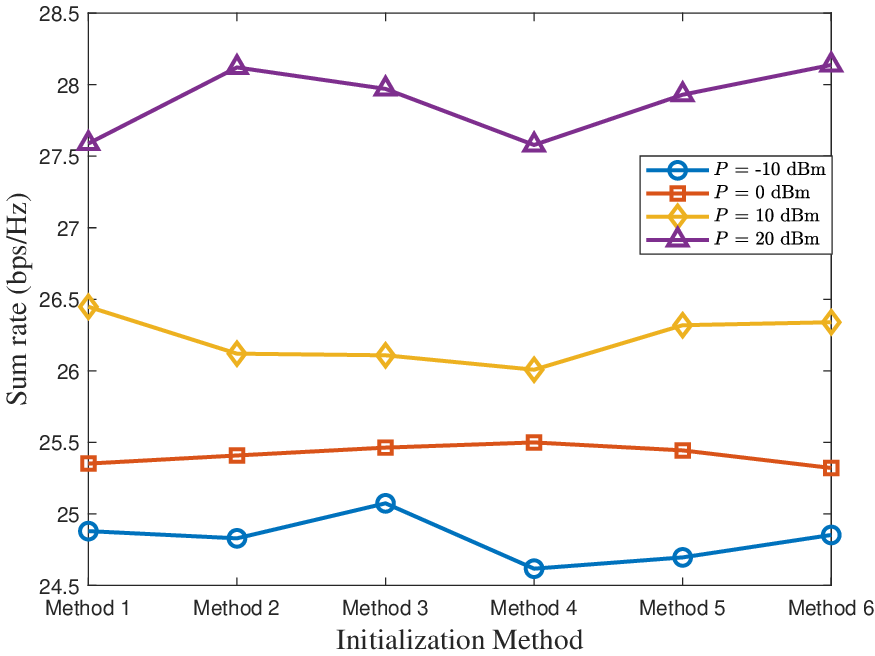} 
    \caption{Sum rate under different initialization strategies.}
    \label{figure10} 
\end{figure}

Under the default multi-waveguide scenario, the subsection evaluates the convergence behavior and robustness of the proposed algorithm. Specifically, Fig. \ref{figure4} illustrates the convergence performance, while Fig. \ref{figure10} examines the impact of different initialization methods on the achieved sum rate.

\subsubsection{Algorithm Convergence Analysis}Fig. \ref{figure4} illustrates the sum rate of the proposed algorithm versus the number of iterations under different carrier frequencies and transmit power. It can be observed that, in all tested scenarios, the system sum rate increases rapidly with the number of iterations and becomes stable after approximately five iterations, demonstrating the fast convergence speed and good numerical stability of the proposed algorithm. In addition, at the 6 GHz carrier frequency, the lower in-waveguide propagation loss enables the PAs to be effectively deployed over a wider range, thereby yielding a higher sum rate. Meanwhile, as the transmit power increases, the achievable system sum rate is further improved, which is consistent with the increase in effective signal power.

\begin{table}[t]
\caption{Initialization Settings for Methods 1--6}
\label{tab:initialization_methods}
\centering
\renewcommand{\arraystretch}{1.1}
\setlength{\tabcolsep}{3.5pt}
\begin{tabular}{|c|p{3.0cm}|p{3.0cm}|}
\hline
\textbf{Method} & \textbf{PA initialization} & \textbf{Power initialization} \\
\hline
Method 1 & Middle-section uniform distribution & Equal allocation \\
\hline
Method 2 & Center-concentrated distribution & Equal allocation \\
\hline
Method 3 & Random distribution & Equal allocation \\
\hline
Method 4 & Middle-section uniform distribution & Random allocation \\
\hline
Method 5 & Center-concentrated distribution & Random allocation \\
\hline
Method 6 & Random distribution & Random allocation \\
\hline
\end{tabular}
\end{table}

\subsubsection{Impact of Different Initialization Strategies}Fig. \ref{figure10} compares the final sum rate performance achieved by the proposed algorithm under six different initialization strategies, whose PA-position and power allocation settings are detailed in Table I. It can be observed that the performance differences among the considered initialization methods remain relatively small under all transmit power levels, indicating that the proposed algorithm is not sensitive to initialization and thus exhibits good robustness. Through iterative updates of the PA positions and power allocation, the algorithm can gradually mitigate the performance disparity caused by different initial points. It is also seen that, although the performance fluctuation among different initialization methods becomes slightly more visible at higher transmit power, the overall gap is still limited, suggesting that the proposed algorithm can still stably converge to similar performance levels.

\subsection{Physical Insights in the Single-Waveguide Scenario}

Different from the default multi-waveguide setting, a single-waveguide scenario is considered to provide physical insights into the underlying system behavior. The spatial dimension parameters are set as $d_1=10$ m, $d_2=10$ m, $D=10$ m, and $h=3$ m. The numbers of waveguides, PAs, and users are set to $M=1$, $N=5$, and $K=2$, respectively. The minimum rate requirement is set to $0.5$ bps/Hz. All sum rate results are averaged over 200 Monte Carlo realizations, and the resolution of the one-dimensional search is set to $G=10000$.

\begin{figure}[t]
    \centering
    \includegraphics[width=3.0in]{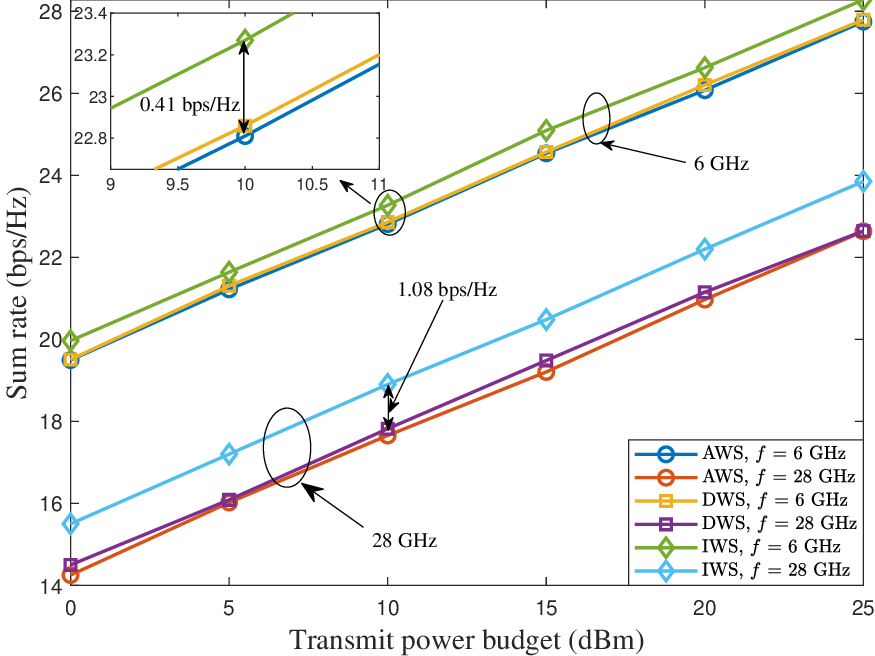} 
    \caption{Sum rate versus the total power budget of PAs at the BS.} 
    \label{figure2} 
\end{figure}

\subsubsection{Effect of In-Waveguide Propagation Loss and Coupling Effect on Sum Rate}Fig. \ref{figure2} illustrates the sum rate versus the total transmit power budget for IWS, DWS, and AWS. As expected, the system sum rate of PASS monotonically increases with the power budget. We can also observe that the performance gap between IWS and DWS expands significantly as the carrier frequency increases, growing from 0.41 bps/Hz at 6 GHz to 1.08 bps/Hz at 28 GHz. The phenomenon is attributed to the fact that the in-waveguide propagation loss increases exponentially with frequency. At higher carrier frequencies, in-waveguide propagation loss is significantly amplified, and its impact on signal transmission can no longer be ignored. Furthermore, the sum rate achieved by AWS is slightly lower than that of DWS, which is caused by coupling effect. The coupling effect necessitates discrete PA positioning along the waveguide, introducing phase misalignment and additional propagation loss that degrades the achievable sum rate. The results highlight that accurate electromagnetic modeling incorporating both propagation loss and coupling effects is essential for reliable performance evaluation and optimization in PASS.

\begin{figure}[t]
    \centering
    \includegraphics[width=3.0in]{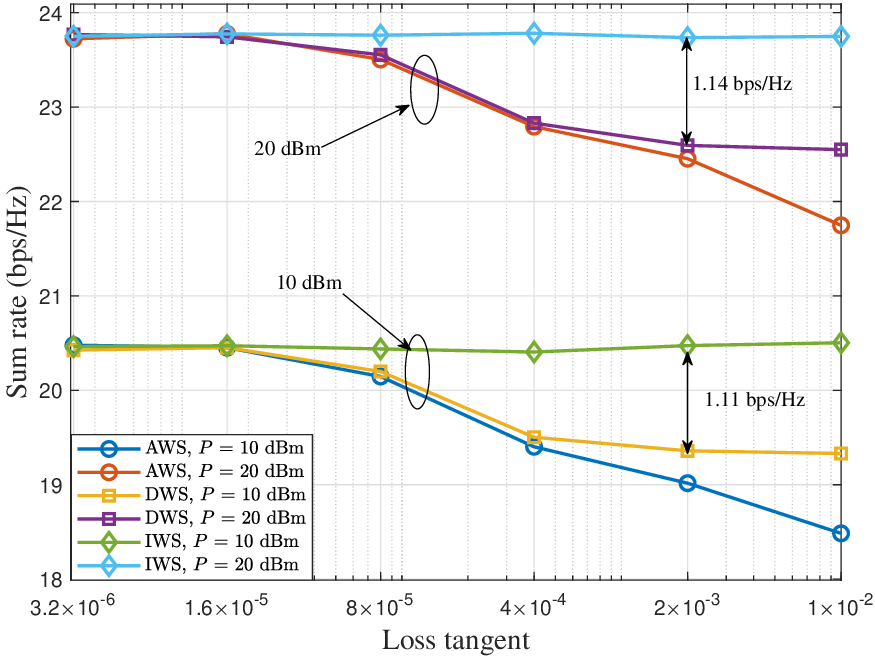} 
    \caption{Sum rate versus the dielectric loss tangent of the waveguide.} 
    \label{figure3} 
\end{figure}

\subsubsection{Impact of dielectric loss tangent on the Sum Rate}Fig. \ref{figure3} shows the system sum rate versus the dielectric loss tangent under different power budgets. The considered dielectric loss tangent values span from an almost lossless to a highly lossy waveguide regime. As the dielectric loss tangent increases, aggravated propagation attenuation causes the sum rate to decrease markedly. When the dielectric loss tangent exceeds 0.002, the performance of DWS no longer degrades as the dielectric loss tangent further increases. The behavior is attributable to the fact that, in this regime, the in-waveguide propagation loss becomes much more severe than the free-space propagation loss. As a result, the optimal PA locations collapse to the feeding point, thereby avoiding additional waveguide propagation. Moreover, a clear performance gap between AWS and DWS emerges once the dielectric loss tangent exceeds 0.0004. The reason is that a minimum spacing must be maintained between adjacent PAs in AWS, which introduces extra in-waveguide propagation distance.

\subsection{Physical Insights in the Multi-Waveguide Scenario}

\begin{figure}[t]
    \centering
    \includegraphics[width=3.0in]{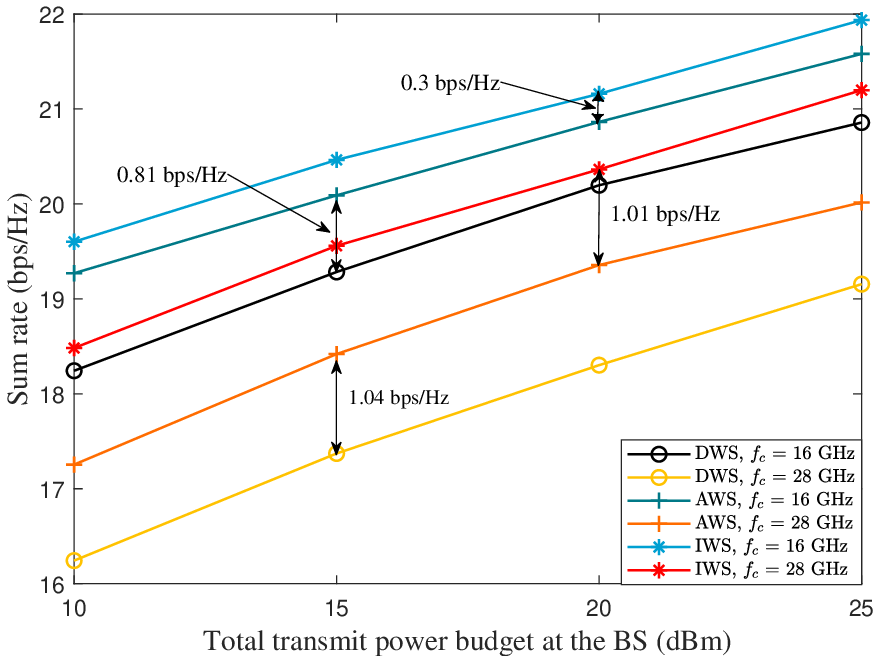} 
    \caption{Sum rate versus the total power budget of PAs at the BS.} 
    \label{figure5} 
\end{figure}

\begin{figure}[t]
    \centering
    \includegraphics[width=3.0in]{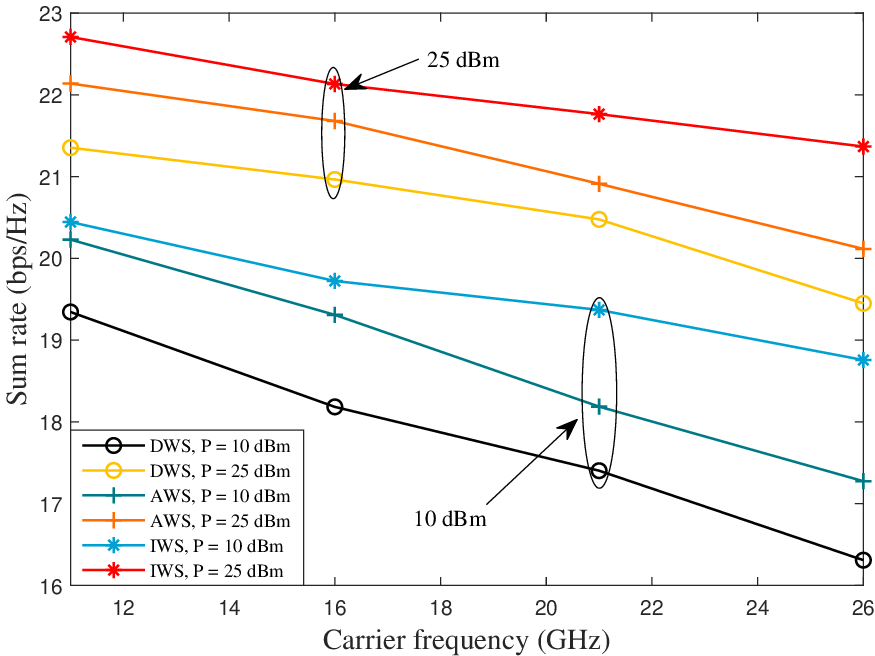} 
    \caption{Sum rate versus the carrier frequency.} 
    \label{figure6} 
\end{figure}

Under the default multi-waveguide scenario, the subsection investigates the impact of physical factors on the system performance. Specifically, Figs. \ref{figure5} and \ref{figure6} show how the key system parameters and underlying physical phenomena influence the achievable sum rate.

\subsubsection{The Impact of Power Budget and Power Redistribution on the Sum Rate}
Fig. \ref{figure5} shows the sum rate versus power budget for multi-waveguide PASS at different carrier frequencies. As the power budget increases, the growth rates of both DWS and AWS gradually slow down. The asymptotic reduction in sum rate growth can be attributed to inter-user interference and the marginal effect of core users’ rates. When the transmit power budget exceeds a certain threshold, the system sum rate is mainly constrained by interference and propagation loss rather than power itself. The results demonstrate that the gains achieved by optimizing resource allocation and user scheduling outweigh those obtained by simply increasing the transmit power budget. Moreover, an unexpected result is observed: AWS achieves a higher sum rate than DWS under the same power budget and carrier frequency. Specifically, at 16 GHz, the sum rate achieved by AWS exceeds that of DWS by 0.81 bps/Hz, while at 28 GHz, AWS outperforms DWS by 1.04 bps/Hz in terms of sum rate. The counterintuitive result arises from the power redistribution gain introduced by discrete PA positioning. To maximize the sum rate, PASS allocates most of the power budget to the core users, who experience the best channel conditions within each time slot, while other users only receive their minimum required rates. Once the core users already achieve high data rates, additional power provides limited improvement in the overall sum rate. Through discrete PA positioning, part of the power budget is reallocated to non-core users, whose performance is significantly improved, thereby enhancing the overall system sum rate.

\subsubsection{The Impact of Carrier Frequency and Propagation Loss on the Sum Rate}
Fig. \ref{figure6} shows the sum rate versus carrier frequency for multi-waveguide PASS under different power budgets. As the carrier frequency increases, both path loss and in-waveguide propagation loss rise, leading to a continuous drop in sum rate. It can be observed that the slpoe of DWS and AWS is significantly smaller at a power budget of 25 dBm than at 10 dBm. The observation is attributed to the fact that a higher power budget more readily satisfies the minimum rate constraint for non-core users. Consequently, PAs can be positioned closer to the core users and allocate more power to serve core users. As a result, the decline in sum rate due to increased frequency is mitigated. In addition, the sum rate gap between AWS and DWS gradually decreases, which is caused by the extra propagation loss introduced by the discrete positioning of PAs. Since in-waveguide propagation loss grows exponentially with carrier frequency, the effect of additional loss becomes significant at high carrier frequencies. As a result, it offsets the gain brought by power redistribution.

\begin{figure}[t]
    \centering
    \includegraphics[width=3.0in]{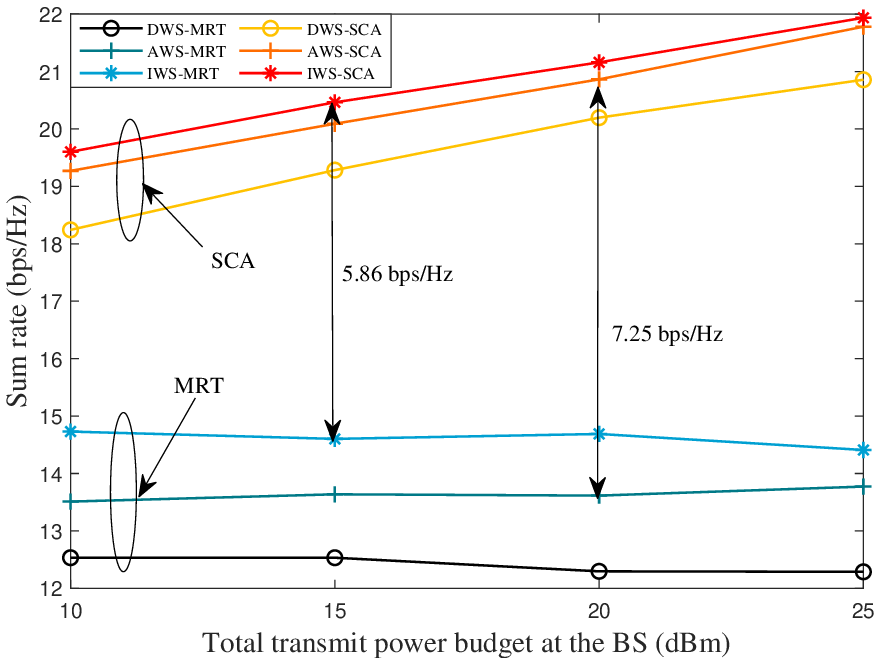} 
    \caption{Sum rate comparison between traditional MRT and SCA.} 
    \label{figure7} 
\end{figure}

\subsection{Performance Comparison and Fairness of the Proposed Scheme}
\begin{figure}[t]
    \centering
    \includegraphics[width=3.0in]{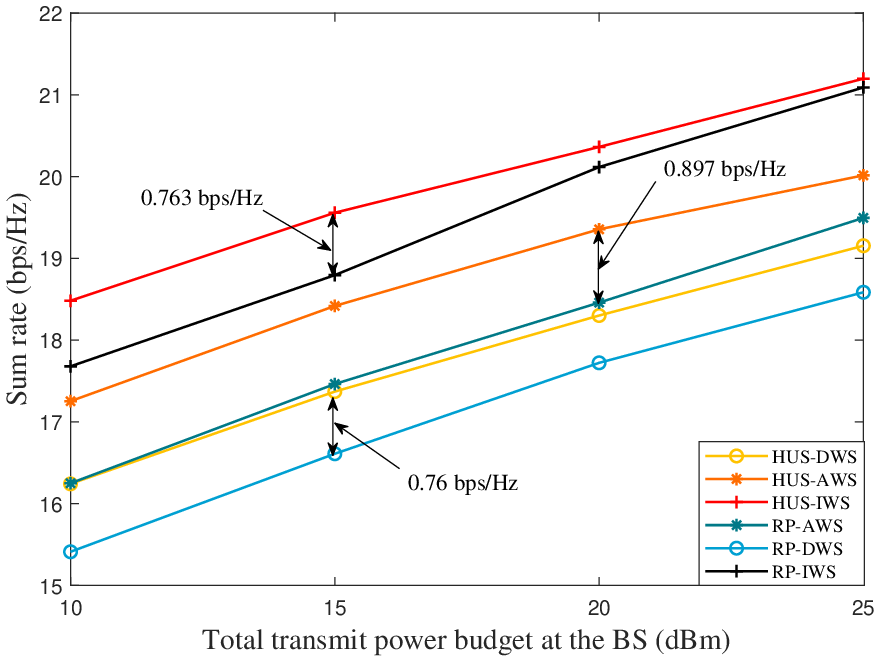} 
    \caption{Sum rate comparison between HUS and random pairing (RP).} 
    \label{figure8} 
\end{figure}

Under the default multi-waveguide scenario, the subsection evaluates the performance superiority and fairness of the proposed scheme. Specifically, Figs. \ref{figure7} and \ref{figure8} compare the proposed scheme with benchmark schemes, while Fig. \ref{figure9} examines its fairness performance.
\subsubsection{Comparison of Power Allocation Algorithm}
Fig. \ref{figure7} shows a sum rate comparison between the proposed power allocation algorithm and traditional maximum ratio transmission (MRT) algorithm at the carrier frequency of $16$ GHz. Across the entire power budget range, the proposed algorithm consistently outperforms MRT, with the performance gap becoming more pronounced as the power budget increases. The advantage originates from the iterative adjustment of power allocation according to the relative positions of users and PAs, which ensures that each user obtains the optimal signal strength while satisfying the minimum rate constraints. The mechanism not only improves signal quality but also mitigates inter-user interference, thereby enhancing the overall transmission efficiency of the system. In contrast, MRT achieves only limited improvement with increasing power budget, since it cannot effectively suppress interference. Although user signal power is maximized, the interference power remains uncontrolled, resulting in severe degradation of the sum rate. The observations confirm the effectiveness of the proposed power allocation algorithm.


\subsubsection{Comparison of User Scheduling Algorithm}

Fig. \ref{figure8} shows the sum rate versus power budget for the HUS algorithm and RP at $28$ GHz. As expected, HUS algorithm achieves a consistently higher sum rate across all power budgets. Specifically, in the schemes considering in-waveguide propagation loss and coupling effects, it brings respective increases in the sum rate of 0.76 bps/Hz and 0.897 bps/Hz. The improvement results from selecting user pairs with low path loss and ensuring spatial separation within each time slot. The strategy leads to more efficient use of power and reduces inter-user interference. HUS algorithm also enables PASS to maintain stable and consistent rate performance. In contrast, RP frequently schedules nearby users, which increases inter-user interference. Additionally, RP is unable to pair the waveguide with the nearest user, which exacerbates the path loss. The two shortcomings lead to a decrease in the sum rate of RP. The results confirm the effectiveness of the HUS algorithm in enhancing the performance of PASS.

\begin{figure}[t]
    \centering
    \includegraphics[width=3.0in]{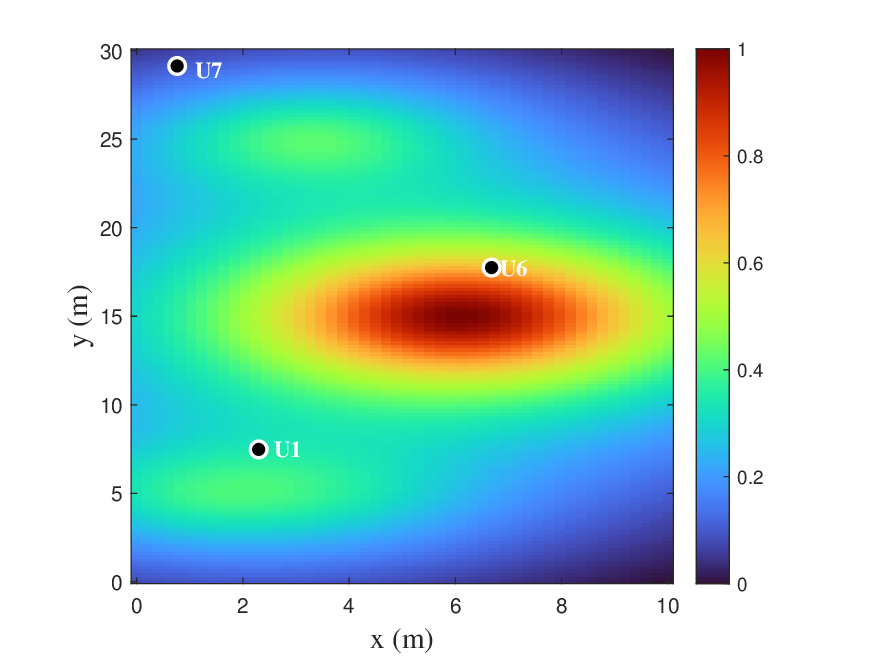} 
    \caption{Heat map of the plane where the users are located at a time slot, with ${R_{\min }} = 1.9$ bps/Hz and $\textit{P} = 20$ dBm.} 
    \label{figure9} 
\end{figure}

\subsubsection{Heat Map Analysis of the Plane Where Users Are Located}
Fig. \ref{figure9} illustrates the heat map of the user plane at a given time slot, where the minimum rate constraint is set to $1.9$ bps/Hz and the power budget is set to $20$ dBm. The x-axis and y-axis are spatial coordinates, and the color bar represents the normalized positioning signal energy. In this time slot, the scheduled users are ${{\bf{U}}_1}$, ${{\bf{U}}_6}$, and ${{\bf{U}}_7}$, whose positions are marked by black dots. It can be clearly observed that the signal energy around ${{\bf{U}}_6}$ is more than twice that around the other users. The reason is that ${{\bf{U}}_6}$ is located closest to the PAs and has the most favorable channel conditions, thereby being regarded as the core user. Under the proposed algorithm, the majority of the power budget is allocated to ${{\bf{U}}_6}$ to maximize the system sum rate, while still ensuring that the minimum rate requirement of $1.9$ bps/Hz is satisfied. The result demonstrates the capability of the proposed algorithm to strike an effective balance between maximizing the sum rate and maintaining user fairness.

\section{Conclusion}

This work focused on the sum rate maximization of multi-waveguide PASS. A physical hardware model was established to jointly characterize in-waveguide propagation loss and coupling effects. Under the TDMA protocol, we proposed a HUS algorithm together with a joint optimization framework. Specifically, the HUS algorithm mitigated path loss by minimizing the sum of squared distances between users and their associated waveguides, while scheduling spatially separated users in each time slot to alleviate inter-user interference. For the joint optimization framework, the PAs’ positions were optimized via one-dimensional search. Power allocation was derived by reformulating the problem through the Lagrangian duality and FP, and then solving it via SCA-based convex optimization. Numerical results indicated that the HUS algorithm efficiently exploited both time domain and space domain resources, thereby reducing path loss and mitigating inter-user interference. The joint optimization framework fully exploited the flexible channel reconfiguration capability of PASS, leading to a substantial improvement in the system sum rate while maintaining fairness among users. In addition, the results highlighted the significant impact of in-waveguide propagation loss and coupling effects on the performance of PASS. In particular, for multi-waveguide PASS, coupling effects induced discrete PA positioning, which redistributed the transmit power and enhanced the system performance. Overall, this study provided a practical and efficient solution for physical modeling, user scheduling, and joint optimization of PASS. The derived insights offered valuable guidance for the design of multi-waveguide transmission in PASS. As a future research direction, the proposed framework can be extended to more general dynamic and coordinated multi-waveguide transmission scenarios, such as adaptive optimization under user mobility, Coordinated Multi-Point, or MIMO beamforming.

\section*{Appendix A: Proof of Theorem 1}

When the functions ${{\mathop{\rm a}\nolimits} _{m,k}}$ and ${{\mathop{\rm b}\nolimits} _{m,k}}$ are fixed, the function about ${{\nu _k}}$ is concavely differentiable, i.e.:
\begin{subequations}\label{eq:mfr}
\begin{align}
&f\left( {{\nu _k}} \right) = \frac{1}{z}{\log _2}\left( {1 + {\nu _k}} \right) - \frac{{{\nu _k}}}{{z\ln 2}} + \frac{{\left( {1 + {\nu _k}} \right){a_{m,k}}}}{{z\ln 2\left( {{a_{m,k}} + {b_{m,k}}} \right)}},\\
&\frac{{{\partial ^2}f\left( {{\nu _k}} \right)}}{{\partial \nu _k^2}} =  - \frac{1}{{z\ln 2}}*\frac{1}{{{{\left( {1 + {\nu _k}} \right)}^2}}} < 0.
\end{align}
\end{subequations}
Thus, the optimal ${\nu _k}^\prime $ can be solved for by making each $\frac{{\delta f\left( {{\nu _k}} \right)}}{{\delta {\nu _k}}}$ equal to zero, i.e.:
\begin{equation}\label{eq:mfr1}
\begin{split}
\frac{1}{{z\ln 2}}*\frac{1}{{1 + {\nu _k}^\prime }} - \frac{1}{{z\ln 2}} + \frac{{{a_{m,k}}}}{{z\ln 2\left( {{a_{m,k}} + {b_{m,k}}} \right)}} = 0.
\end{split}
\end{equation}

The optimal solution ${\nu _k}^\prime  = \frac{{{{\mathop{\rm a}\nolimits} _{m,k}}}}{{{{\mathop{\rm b}\nolimits} _{m,k}}}}$ can be solved. The optimal solution ${\nu _k}^\prime$ is brought into the function $f\left( {{\nu _k}} \right)$,which can be obtained as follows:
\begin{equation}\label{eq:mfr2}
\begin{split}
f\left( {{\nu _k}^\prime } \right) = \frac{1}{z}{\log _2}\left( {1 + \frac{{{{\mathop{\rm a}\nolimits} _{m,k}}}}{{{{\mathop{\rm b}\nolimits} _{m,k}}}}} \right).
\end{split}
\end{equation}
Therefore, the proof is complete.

\section*{Appendix B: Proof of Theorem 2}

When the functions ${{\mathop{\rm a}\nolimits} _{m,k}}$ and ${{\mathop{\rm b}\nolimits} _{m,k}}$ are fixed, the function about ${{y_k}}$ is given by:
\begin{equation}\label{eq:mfy}
\begin{split}
f\left( {{{y_k}}} \right) = 2{{y_k}}\sqrt {{{\mathop{\rm a}\nolimits} _{m,k}}}  - {y_k^2}\left( {{{\mathop{\rm a}\nolimits} _{m,k}} + {{\mathop{\rm b}\nolimits} _{m,k}}} \right).
\end{split}
\end{equation}
Because ${{{\mathop{\rm a}\nolimits} _{m,k}}} > 0$ and ${{{\mathop{\rm b}\nolimits} _{m,k}}} > 0$, the second order derivative of the function $f\left( {{y_k}} \right)$ is constantly less than zero:
\begin{equation}\label{eq:mfy2}
\begin{split}
\frac{{{\partial ^2}f\left( {{y_k}} \right)}}{{\partial y_k^2}} =  - 2\left( {{{\mathop{\rm a}\nolimits} _{m,k}} + {{\mathop{\rm b}\nolimits} _{m,k}}} \right) < 0.
\end{split}
\end{equation}
Then, the optimal solution can be solved by derivation:
\begin{equation}\label{eq:mfy1}
\begin{split}
2\sqrt {{{\mathop{\rm a}\nolimits} _{m,k}}}  - 2{{{y_k}}^\prime}\left( {{{\mathop{\rm a}\nolimits} _{m,k}} + {{\mathop{\rm b}\nolimits} _{m,k}}} \right) = 0.
\end{split}
\end{equation}

Clearly, \eqref{eq:mfy1} can be solved for ${{y_k}}^\prime  = \frac{{\sqrt {{{\mathop{\rm a}\nolimits} _{m,k}}} }}{{{{\mathop{\rm a}\nolimits} _{m,k}} + {{\mathop{\rm b}\nolimits} _{m,k}}}}$. The optimal solution is brought into the function:
\begin{equation}\label{eq:mfy3}
\begin{split}
f\left( {{{y_k}}^\prime } \right) = \frac{{{{\mathop{\rm a}\nolimits} _{m,k}}}}{{{{\mathop{\rm a}\nolimits} _{m,k}} + {{\mathop{\rm b}\nolimits} _{m,k}}}}.
\end{split}
\end{equation}
Therefore, the proof is complete.

\bibliographystyle{IEEEtran}
\bibliography{PA.bib}

@ARTICLE{6375940,
  author={Rusek, Fredrik and Persson, Daniel and Lau, Buon Kiong and Larsson, Erik G. and Marzetta, Thomas L. and Edfors, Ove and Tufvesson, Fredrik},
  journal={IEEE Signal Processing Mag.}, 
  title={{Scaling Up MIMO: Opportunities and Challenges with Very Large Arrays}}, 
  year={2013},
  volume={30},
  number={1},
  pages={40-60},
  month={Dec.},
  doi={10.1109/MSP.2011.2178495}}

@article{larsson2014massive,
  title={{Massive MIMO for Next Generation Wireless Systems}},
  author={Larsson, Erik G and Edfors, Ove and Tufvesson, Fredrik and Marzetta, Thomas L},
  journal={IEEE Commun. Mag.},
  volume={52},
  number={2},
  pages={186--195},
  year={2014},
  month={Feb.},
  publisher={IEEE}
}

@ARTICLE{8869705,
  author={Saad, Walid and Bennis, Mehdi and Chen, Mingzhe},
  journal={IEEE Network}, 
  title={{A Vision of 6G Wireless Systems: Applications, Trends, Technologies, and Open Research Problems}}, 
  year={2020},
  volume={34},
  number={3},
  pages={134-142},
  month={Oct.},
  doi={10.1109/MNET.001.1900287}}

@article{ahmadzadeh2025enhancedovertheairfederatedlearning,
      title={{Enhanced Over-the-Air Federated Learning Using AI-based Fluid Antenna System}}, 
      author={Mohsen Ahmadzadeh and Saeid Pakravan and Ghosheh Abed Hodtani and Ming Zeng and Jean-Yves Chouinard and Leslie A. Rusch},
      year={2025},
      journal={arXiv},
      volume={2407.03481},
}

@INPROCEEDINGS{11100909,
  author={Pakravan, Saeid and Ahmadzadeh, Mohsen and Zeng, Ming and Hodtani, Ghosheh Abed and Chouinard, Jean-Yves and Rusch, Leslie A},
  booktitle={2024 Proc. IEEE Globecom Workshops}, 
  title={{Robust Resource Allocation for Over-The-Air Computation Networks with Fluid Antenna Array}}, 
  year={2024},
  pages={1-6},
  month={Dec.},
  doi={10.1109/GCWkshp64532.2024.11100909}}

@INPROCEEDINGS{10437926,
  author={Wu, Yifei and Xu, Dongfang and Ng, Derrick Wing Kwan and Gerstacker, Wolfgang and Schober, Robert},
  booktitle={2023 IEEE Global Commun. Conf.}, 
  title={{Movable Antenna-Enhanced Multiuser Communication: Jointly Optimal Discrete Antenna Positioning and Beamforming}}, 
  year={2023},
  pages={7508-7513},
  month={Feb.},
  doi={10.1109/GLOBECOM54140.2023.10437926}}

@ARTICLE{9424177,
  author={Liu, Yuanwei and Liu, Xiao and Mu, Xidong and Hou, Tianwei and Xu, Jiaqi and Di Renzo, Marco and Al-Dhahir, Naofal},
  journal={IEEE Commun. Surveys \& Tuts.}, 
  title={{Reconfigurable Intelligent Surfaces: Principles and Opportunities}}, 
  year={2021},
  volume={23},
  number={3},
  pages={1546-1577},
month={May},
  doi={10.1109/COMST.2021.3077737}}

@ARTICLE{9427474,
  author={Zhang, Shuowen and Zhang, Rui},
  journal={IEEE Trans. Commun.}, 
  title={{Intelligent Reflecting Surface Aided Multi-User Communication: Capacity Region and Deployment Strategy}}, 
  year={2021},
  volume={69},
  number={9},
  month={May},
  pages={5790-5806},
  doi={10.1109/TCOMM.2021.3079128}}

@ARTICLE{10909572,
  author={Zhu, Lipeng and Ma, Wenyan and Xiao, Zhenyu and Zhang, Rui},
  journal={IEEE Trans. Commun.}, 
  title={{Movable Antenna Enabled Near-Field Communications: Channel Modeling and Performance Optimization}}, 
  year={2025},
  volume={73},
  number={9},
  pages={7240-7256},
  doi={10.1109/TCOMM.2025.3547783}}

@ARTICLE{10643473,
  author={Ma, Wenyan and Zhu, Lipeng and Zhang, Rui},
  journal={IEEE Trans. Wireless Commun.}, 
  title={{Movable Antenna Enhanced Wireless Sensing via Antenna Position Optimization}}, 
  year={2024},
  volume={23},
  number={11},
  pages={16575-16589},
  doi={10.1109/TWC.2024.3443293}}

@ARTICLE{9690635,
  author={Yu, Xianghao and Jamali, Vahid and Xu, Dongfang and Ng, Derrick Wing Kwan and Schober, Robert},
  journal={IEEE Wireless Commun.}, 
  title={{Smart and Reconfigurable Wireless Communications: From IRS Modeling to Algorithm Design}}, 
  year={2021},
  volume={28},
  number={6},
  pages={118-125},
    month={Jan.},
  doi={10.1109/MWC.001.2100145}}

@ARTICLE{9140329,
  author={Di Renzo, Marco and Zappone, Alessio and Debbah, Merouane and Alouini, Mohamed-Slim and Yuen, Chau and de Rosny, Julien and Tretyakov, Sergei},
  journal={IEEE J. Sel. Areas Commun.}, 
  title={{Smart Radio Environments Empowered by Reconfigurable Intelligent Surfaces: How It Works, State of Research, and The Road Ahead}}, 
  year={2020},
  volume={38},
  number={11},
  pages={2450-2525},
  month={Nov.},
  doi={10.1109/JSAC.2020.3007211}}

@ARTICLE{10286328,
  author={Zhu, Lipeng and Ma, Wenyan and Zhang, Rui},
  journal={IEEE Commun. Mag.}, 
  title={{Movable Antennas for Wireless Communication: Opportunities and Challenges}}, 
  year={2024},
  volume={62},
  number={6},
  pages={114--120},
  month={jun},
  doi={10.1109/MCOM.001.2300212}}

@article{suzuki2022pinching,
  title={{Pinching antenna: Using a dielectric waveguide as an antenna}},
  author={Suzuki, Hiroshi Okazaki Yasunori and Kawai, Kunihiro},
  journal={NTT DOCOMO Technical J},
  volume={23},
  number={3},
  pages={5--12},
  month={Jan.},
  year={2022}

}

@article{yang2025pinchingantennasprinciplesapplications,
      title={{Pinching Antennas: Principles, Applications and Challenges}}, 
      author={Zheng Yang and Ning Wang and Yanshi Sun and Zhiguo Ding and Robert Schober and George K. Karagiannidis and Vincent W. S. Wong and Octavia A. Dobre},
      year={2025},
       journal={arXiv},
      volume={2501.10753}
}

@ARTICLE{10909665,
  author={Tegos, Sotiris A. and Diamantoulakis, Panagiotis D. and Ding, Zhiguo and Karagiannidis, George K.},
  journal={IEEE Wireless Commun. Lett.}, 
  title={{Minimum Data Rate Maximization for Uplink Pinching-Antenna Systems}}, 
  year={2025},
  volume={14},
  number={5},
  pages={1516-1520},
   month={May},
  doi={10.1109/LWC.2025.3547956}}

@article{liu2025pinching,
  title={{Pinching-Antenna Systems (PASS): Architecture Designs, Opportunities, and Outlook}}, 
  author={Liu, Yuanwei and Wang, Zhaolin and Mu, Xidong and Ouyang, Chongjun and Xu, Xiaoxia and Ding, Zhiguo},
  year={2025},
       journal={arXiv},
      volume={2501.18409}
}

@article{liu2025pinchingantennasystemspasstutorial,
      title={{Pinching-Antenna Systems (PASS): A Tutorial}}, 
      author={Yuanwei Liu and Hao Jiang and Xiaoxia Xu and Zhaolin Wang and Jia Guo and Chongjun Ouyang and Xidong Mu and Zhiguo Ding and Arumugam Nallanathan and George K. Karagiannidis and Robert Schober},
      year={2025},
       journal={arXiv},
      volume={2508.07572}, 
}

@article{ding2024flexibleantennasystemspinchingantennaperspective,
      title={Flexible-Antenna Systems: A Pinching-Antenna Perspective}, 
      author={Zhiguo Ding and Robert Schober and H. Vincent Poor},
      year={2024},
       journal={arXiv},
      volume={2412.02376}, 
}

@article{xu2025jointtransmitpinchingbeamforming,
      title={{Joint Transmit and Pinching Beamforming for Pinching Antenna Systems (PASS): Optimization-Based or Learning-Based?}}, 
      author={Xiaoxia Xu and Xidong Mu and Yuanwei Liu and Arumugam Nallanathan},
      year={2025},
       journal={arXiv},
      volume={2502.08637}
}

@article{ouyang2025arraygainpinchingantennasystems,
      title={{Array Gain for Pinching-Antenna Systems (PASS)}}, 
      author={Chongjun Ouyang and Zhaolin Wang and Yuanwei Liu and Zhiguo Ding},
      year={2025},
      journal={arXiv},
      volume={2501.05657}
}

@article{wang2024antennaactivationnomaassisted,
      title={{Antenna Activation for NOMA Assisted Pinching-Antenna Systems}}, 
      author={Kaidi Wang and Zhiguo Ding and Robert Schober},
      year={2024},
       journal={arXiv},
      volume={2412.13969} 
}

@ARTICLE{11195810,
  author={Hou, Tianwei and Liu, Yuanwei and Nallanathan, Arumugam},
  journal={IEEE Trans. Commun.}, 
  title={{On the Performance of Uplink Pinching Antenna Systems (PASS)}}, 
  year={2025},
  volume={},
  number={},
  pages={1-1},
  doi={10.1109/TCOMM.2025.3618726}}

@article{zhang2025uplinksumratemaximization,
      title={{Uplink Sum Rate Maximization for Pinching Antenna-Assisted Multiuser MISO}}, 
      author={Jiarui Zhang and Hao Xu and Chongjun Ouyang and Qiuyun Zou and Hongwen Yang},
      year={2025},
      journal={arXiv},
      volume={2504.16577} 
}

@article{bereyhi2025mimopassuplinkdownlinktransmission,
      title={{MIMO-PASS: Uplink and Downlink Transmission via MIMO Pinching-Antenna Systems}}, 
      author={Ali Bereyhi and Chongjun Ouyang and Saba Asaad and Zhiguo Ding and H. Vincent Poor},
      year={2025},
      journal={arXiv},
      volume={2503.03117} 
}

@article{tyrovolas2025performanceanalysispinchingantennasystems,
      title={{Performance Analysis of Pinching-Antenna Systems}}, 
      author={Dimitrios Tyrovolas and Sotiris A. Tegos and Panagiotis D. Diamantoulakis and Sotiris Ioannidis and Christos K. Liaskos and George K. Karagiannidis},
      year={2025},
      journal={arXiv},
      volume={2502.06701}
}

@article{hu2025sumratemaximizationpinchingantennaassisted,
      title={{Sum-Rate Maximization for Pinching Antenna-assisted NOMA Systems with Multiple Dielectric Waveguides}}, 
      author={Shaokang Hu and Ruotong Zhao and Yihuan Liao and Derrick Wing Kwan Ng and Jinhong Yuan},
      year={2025},
      journal={arXiv},
      volume={2503.10060}
}

@article{li2025pinchingantennaaidedwirelesspowered,
      title={{Pinching Antenna-Aided Wireless Powered Communication Networks}}, 
      author={Yixuan Li and Hongbo Xu and Ming Zeng and Yuanwei Liu},
      year={2025},
      journal={arXiv},
      volume={2506.00355}
}

@article{sun2025multiuserbeamformingpinchingantennasystems,
      title={{Multiuser Beamforming for Pinching-Antenna Systems: An Element-wise Optimization Framework}}, 
      author={Mingjun Sun and Chongjun Ouyang and Shaochuan Wu and Yuanwei Liu},
      year={2025},
      journal={arXiv},
      volume={2506.03770}
}

@article{zhao2025pinchingantennasystemsenabledmultiusercommunications,
      title={{Pinching-Antenna Systems-Enabled Multi-User Communications: Transmission Structures and Beamforming Optimization}}, 
      author={Jingjing Zhao and Haowen Song and Xidong Mu and Kaiquan Cai and Yanbo Zhu and Yuanwei Liu},
      year={2025},
      journal={arXiv},
      volume={2508.14458}
}

@article{xu2025pinchingantennasystemspasspower,
      title={{Pinching-Antenna Systems (PASS): Power Radiation Model and Optimal Beamforming Design}}, 
      author={Xiaoxia Xu and Xidong Mu and Zhaolin Wang and Yuanwei Liu and Arumugam Nallanathan},
      year={2025},
      journal={arXiv},
      volume={2505.00218}
}

@article{wang2025modelingbeamformingoptimizationpinchingantenna,
      title={{Modeling and Beamforming Optimization for Pinching-Antenna Systems}}, 
      author={Zhaolin Wang and Chongjun Ouyang and Xidong Mu and Yuanwei Liu and Zhiguo Ding},
      year={2025},
      journal={arXiv},
      volume={2502.05917}
}

@article{zhao2025waveguidedivisionmultipleaccess,
      title={{Waveguide Division Multiple Access for Pinching-Antenna Systems (PASS)}}, 
      author={Jingjing Zhao and Xidong Mu and Kaiquan Cai and Yanbo Zhu and Yuanwei Liu},
      year={2025},
      journal={arXiv},
      volume={2502.17781} 
}

@article{wang2025antennaactivationresourceallocation,
      title={{Antenna Activation and Resource Allocation in Multi-Waveguide Pinching-Antenna Systems}}, 
      author={Kaidi Wang and Zhiguo Ding and George K. Karagiannidis},
      year={2025},
      journal={arXiv},
      volume={2505.02864} 
}

@book{pozar2021microwave,
  title={{Microwave Engineering}},
  author={Pozar, David M},
  year={2011},
  publisher={John Wiley \& Sons},
  address={Hoboken, NJ, USA},
  edition = {4th}
}

@book{okamoto2021fundamentals,
  title={{Fundamentals of Optical Waveguides}},
  author={Okamoto, Katsunari},
  year={2021},
  publisher={Elsevier}
}

@ARTICLE{9738442,
  author={Zhang, Haiyang and Shlezinger, Nir and Guidi, Francesco and Dardari, Davide and Imani, Mohammadreza F. and Eldar, Yonina C.},
  journal={IEEE Trans. Wireless Commun.}, 
  title={{Beam Focusing for Near-Field Multiuser MIMO Communications}}, 
  year={2022},
  volume={21},
  number={9},
  pages={7476-7490},
  doi={10.1109/TWC.2022.3158894},
  month={Sep.},}

@ARTICLE{8310563,
  author={Shen, Kaiming and Yu, Wei},
  journal={IEEE Trans. Signal Process.}, 
  title={{Fractional Programming for Communication Systems—Part II: Uplink Scheduling via Matching}}, 
  year={2018},
  volume={66},
  number={10},
  pages={2631-2644},
  doi={10.1109/TSP.2018.2812748},
    month={May},}

@ARTICLE{8314727,
  author={Shen, Kaiming and Yu, Wei},
  journal={IEEE Trans. Signal Process.}, 
  title={{Fractional Programming for Communication Systems—Part I: Power Control and Beamforming}}, 
  year={2018},
  volume={66},
  number={10},
  pages={2616-2630},
  doi={10.1109/TSP.2018.2812733},
  ISSN={1941-0476},
  month={May},}

\end{document}